\documentclass[qsecnum,amsmath,preprintnumbers,superscriptaddress,nofootinbib,aps,prd,10pt,a4paper]{revtex4-1}
\pdfoutput=1
\usepackage[utf8]{inputenc}
\usepackage{amsmath}
\usepackage{amsfonts}
\usepackage{amssymb}
\usepackage{amsthm}
\usepackage[colorlinks=black, citecolor=green, linkcolor=blue, linktocpage=true]{hyperref}
\usepackage{xcolor}
\usepackage{bm}
\usepackage{graphicx}
\usepackage{feynmp-auto}
\usepackage{comment}
\usepackage{mathrsfs}
\usepackage{caption}
\usepackage{subcaption}
\usepackage{framed}
\def\baselinestretch{1.2}
 \def\be{\begin{equation}}
\def\ee{\end{equation}}
 \def\ba{\begin{align}}
\def\ea{\end{align}}
\def\bea{\begin{eqnarray}}
\def\eea{\end{eqnarray}}

\def\a{\alpha}

\def\m{\mu}
\def\n{\nu}

\newcommand{\bseq}{\begin{subequations}}
\newcommand{\eseq}{\end{subequations}}

\begin{document}
\title{{\bf Hawking Radiation in Lorentz Violating Gravity: A Tale of Two Horizons}}

\author{F. Del Porro}
\email[]{fdelporr@sissa.it}
\address{SISSA, Via Bonomea 265, 34136 Trieste, Italy and INFN Sezione di Trieste}
\address{IFPU - Institute for Fundamental Physics of the Universe, Via Beirut 2, 34014 Trieste, Italy}

\author{M. Herrero-Valea}
\email[]{mherrero@ifae.es}
\address{Institut de Fisica d’Altes Energies (IFAE), The Barcelona Institute of Science and Technology, Campus UAB, 08193 Bellaterra (Barcelona) Spain}

\author{S. Liberati}
\email[]{liberati@sissa.it}

\author{M. Schneider}
\email[]{mschneid@sissa.it}

\address{SISSA, Via Bonomea 265, 34136 Trieste, Italy and INFN Sezione di Trieste}
\address{IFPU - Institute for Fundamental Physics of the Universe, Via Beirut 2, 34014 Trieste, Italy}

\date{\today}

\begin{abstract}

Since their proposal, Lorentz violating theories of gravity have posed a potential threat to black hole thermodynamics, as superluminal signals appeared to be incompatible with the very black hole notion.
Remarkably, it was soon realized that in such theories causally disconnected regions of space-time can still exist thanks to the presence of universal horizons: causal barriers for signals of arbitrary high speed. Several investigations, sometimes with contrasting results, have been performed so to determine if these horizons can be associated with healthy thermodynamic properties similar to those associated with Killing horizons in General Relativity. In this work we offer what we deem to be the final picture emerging from this and previous studies. In summary we show that: 1) there is a thermal, and most of all species-independent, emission associated to universal horizons, determined by their surface gravity; 2) due to the modified dispersion relation of the matter fields, the low energy part of the emitted spectrum is affected by the presence of the Killing horizon, in a way similar to an effective refractive index, leading at low energies (w.r.t. the Lorentz breaking scale) to an emission that mimics a standard Hawking spectrum (i.e.~one determined by the Killing horizon surface gravity); 3) the whole picture is compatible with a globally well defined vacuum state i.e. an Unruh state associated with preferred observers, which however at very low energies it is basically indistinguishable from the standard Unruh vacuum associated to metric free-falling observers. One can then conclude that Hawking radiation is remarkably resilient even within the context of gravitational theories entailing the breakdown of local Lorentz invariance. 

\end{abstract}

\maketitle
{\renewcommand{\baselinestretch}{1.2} \parskip=0pt
\setcounter{tocdepth}{2}
\tableofcontents}
\newpage

\section{Introduction}

The fact that black holes radiate is one of the most unexpected, albeit also most studied, implications of Quantum Field Theory (QFT) in curved space-times \cite{Hawking75}. Event horizons, representing causal boundaries for the propagation of classical fields, behave as black bodies, emitting a thermal spectrum of particles with temperature given by the Bekenstein-Hawking formula $T=(8\pi M)^{-1}$, where $M$ is the black hole mass. This result holds in General Relativity (GR), as well as in most models of modified gravity, since it can be linked to local properties of the space-time geometry, dictated by covariance and diffeomorphism invariance.

This conclusion can however be contested in the presence of Lorentz violating gravity, where local Lorentz invariance (LLI) is abandoned in first place by introducing a unit-norm, time-like dynamical vector field $U^\m$, the \emph{aether}. Its presence changes the local causal structure of space-time, allowing for superluminal propagation up to arbitrary speeds, so that the usual definition of a trapping horizon loses its meaning. This can be in principle worrisome for several reasons. First, the absence of a sensible horizon allows observers sitting far from the gravitational well of the black hole to probe distances arbitrarily close to the central singularity, so violating the (weak) cosmic censorship conjecture. Second, the emission of Hawking radiation in a thermal ensemble by the horizon is intimately related to the thermodynamical aspects of black holes: losing the relativistic notion of horizons, one could naturally wonder whether black hole thermodynamics can survive in this situation, and indeed early investigations pointed out a possible breakdown of the second law of thermodynamics in these settings, and the possibility to build {\em perpetuum mobiles}~\cite{Dubovsky:2006vk,Eling:2007qd,Jacobson:2010fat}. 

However, it was soon realized that these early investigations were incomplete, in the sense that they studied the motion of matter endowed with Lorentz violating dispersion relations but neglected to consider the full impact of the aether in these geometries. Its mere presence introduces a preferred threading which provides a notion of universal causality for all observers, and replaces the standard light-cone by the concept of causal and a-causal curves with respect to the aether flow \cite{Bhattacharyya16}. This can be described in full generality (at least to second order in derivatives) by Einstein-Aether (EA) gravity~\cite{Jacobson:2000xp}, where the aether couples dynamically to the Einstein-Hilbert action. From an effective field theory perspective, it encodes the low energy limit of any Lorentz violating theory of gravitation, and in particular, of Ho\v rava gravity \cite{Horava09}, which will be our focus here, as we will discuss below.

Furthermore, in spherically symmetric and static solutions, $U^\m$ is constrained to be hypersurface orthogonal, defining a preferred foliation in space-like co-dimension one hypersufaces. Remarkably, it was then realized that when the aether flow becomes orthogonal to the time-like Killing vector $\chi^\m$ at a certain radial surface, this corresponds to a leaf of the foliation which becomes a compact causal boundary for every motion, trapping any worldline on its interior, and is hence named \emph{universal horizon} (UH) \cite{Berglund12}. Its existence allows in principle to bypass the obstructions previously described -- it protects exterior observers from probing the area neighboring the singularity, and it is a strong candidate to replace the Killing horizon in behaving as a black body with a fixed temperature.  

The thermal properties of the universal horizon have been hence extensively studied with several techniques in the past decade, including the tunneling method \cite{DelPorro22_tunneling,Berglund13}, and the description of the dynamics of a collapsing shell \cite{Herrero-Valea20}. Note however that early works \cite{Herrero-Valea20,DelPorro22,DelPorro22_tunneling} -- including some by the present authors -- claimed a non-universal character of the particle emission, with a temperature depending on the behavior of the emitted fields in the UV. This in turn would imply a possible, very dangerous ``species problem", at least without appealing to some universal of behaviour of matter in the far UV~\cite{Herrero-Valea20,Ding:2016srk}. We argued in a more recent paper~\cite{Schneider23}, that this is not the case by showing that a rigorous calculation in the tunneling approach requires the non-universal (species dependent) terms in the Hawking radiation derivation to cancel against previously ignored sub-leading corrections. The resulting temperature of the UH is hence \emph{universal}, and independent on the nature of the particle species. In what follows, we shall come back to this point in greater detail.

However, this is not the end of the story. Even though the UH radiates thermally, the shape of the radiation arriving to an observer sitting far from the gravitational well might be very different from a thermal spectrum. Indeed, due to the interaction with the aether, rays of arbitrary energy do not follow geodesics within these spacetimes, but instead their trajectory depends on their Killing energy. Rays with large energies will travel almost unaltered from the UH to the asymptotic region. However, those with small energies will behave similarly to their analogues in a general relativistic theory. In particular, they are influenced by the presence of the Killing horizon, and they end up lingering in its neighborhood in an energy dependent way \cite{Cropp14}. This imprints an effect in the observed spectrum at large radii, which will not be exactly thermal anymore.

In this paper we offer a complete treatment of the particle production by Lorentz-violating black holes, from the description of the thermal emission at the UH, to a full understanding of the influence of the Killing horizon on the spectrum observed at large radii. We will do this by providing a careful and complete analysis of the space of solutions of a Lorentz-violating scalar field of the Lifshitz kind, within the WKB approximation; and by studying the production and subsequent propagation of radiation up to the asymptotic region. More in detail, in Section \ref{s:AEG} we review our geometrical background, describing static spherically symmetric black holes in EA gravity, and the equations describing matter propagation in the presence of a Lorentz violating coupling. In section \ref{sec:MattKin} we then move to describe the propagating matter modes, giving a full picture of the solutions of the field equations within the WKB approximation. Then, in Section \ref{s:PP} we derive the thermal properties associated to the chosen black hole solutions. With a tunneling approach, we analyze both the universal and the Killing horizon, showing how particles behave in their neighborhoods. These results lead naturally to Section \ref{s:QS}, where we discuss the shape of a possible vacuum state compatible with our result, and which encodes the thermal properties derived previously. Together with this, we will analyze the trustfulness of the WKB ansatz that we have used in all the previous sections. Hereinafter, we work with the mostly-minus sign convention for the metric. 

\section{Lorentz Violating Gravity}\label{s:AEG}

Throughout this analysis, we will focus on a gravitational theory endowed with a preferred physical frame. In particular, and motivated by a quantum gravity perspective, we consider Ho\v rava gravity \cite{Horava09} -- see \cite{Herrero-Valea:2023zex} for a recent review. By construction, space-times in Ho\v rava gravity are composed out of foliated manifolds, that admit \textit{a priori} an anisotropic scaling between time $\tau$ and space $x^i$ at high energies
\begin{align}\label{eq:Lifshitz_scaling}
    \tau\rightarrow \hat{\varrho}^{n} \tau,\quad x^i\rightarrow \hat{\varrho} x^i \quad\mbox{with}\quad\hat\varrho\in\mathbb{R}\,,
\end{align}
in order to guarantee a UV-regular behavior of the theory, leading to absence of ghosts and power-counting renormalizability \cite{Horava09,Blas10}, or even to full renormalizability in some cases \cite{Barvinsky:2015kil}. Moreover, the Lifshitz scaling \eqref{eq:Lifshitz_scaling} implies a notion of absolute time, that naturally factorizes the manifold $\cal M$ as $\mathcal{M}\simeq\mathbb{R}\times\Sigma$, where the co-dimension one submanifold $\Sigma$ represents a constant time leaf. Geometrically, this is similar to introducing a vector field $U^\mu$ that is normalized and globally timelike. Since $\mathcal{M}$ is foliated along $U$, the vector field is orthogonal to $\Sigma$ and thus defines the time direction. In a phenomenological perspective we shall focus on the low-energy limit of Ho\v rava gravity, the so-called \emph{khronometric gravity} \cite{Blas:2010hb}, corresponding to a gravitational theory quadratic in derivatives, and with a hypersurface orthogonal timelike vector field. Its action can be described through that of EA gravity \cite{Jacobson01} by imposing hypersurface orthogonality on the aether \cite{Herrero-Valea:2023zex}. In the remainder, we shall focus on this model.

\subsection{Lorentz-violating Black Holes}
Within the geometrical setup just described, we focus on spherically symmetric and static solutions exhibiting a Killing horizon (KH), generically described by the line-element
\begin{align}\label{eq:metric}
{\rm d}s^2=F(r){\rm d}t^2 -\frac{B(r)^2}{F(r)}{\rm d}r^2 -r^2 {\rm d}\mathbb{S}_2,  
\end{align}
where $\mathbb{S}_2$ denotes the two-sphere. The corresponding aether, reflecting the symmetries of space-time, is given by
\begin{align}\label{eq:aether}
   U_\m {\rm d}x^\m=\frac{1+F(r)A(r)^2}{2 A(r)}{\rm d}t +\frac{B(r)}{2A(r)}\left(\frac{1}{F(r)}-A(r)^2\right){\rm d}r.
\end{align}
Here $F(r)$, $B(r)$, and $A(r)$ are arbitrary functions to be determined by the EA field equations, while the normalization of the aether is chosen for convenience, and automatically satisfies the unit norm condition $U_\mu U^\mu=1$. Notice that this solution admits a Killing vector $\chi_\m {\rm d}x^\m=F(r){\rm d}t$, with squared norm $|\chi|^2=F(r)$. Hence the solution of $F(r)=0$ signals the position of the KH. Noticeably, static, spherically symmetric space-times automatically enforce hypersurface orthogonality onto $U^\m$ without any need of extra constraints. Thus, it can always be written as the (normalized) gradient of a scalar field, dubbed khronon, which describes the foliation into space-like leafs $\Sigma$ orthogonal to $U^\m$.

From a geometrical point of view, the foliation endows the geometry with a time-orientation, which becomes here a fundamental building block of space-time. By construction, the manifold becomes a time-ordered stack of spatial hypersurfaces, and loses full diffeomorphism invariance. Instead, all solutions to Ho\v rava gravity obey the subgroup of foliation preserving diffeomorphisms (FDiff)
\begin{align}\label{eq:FDiff}
    \tau \rightarrow \tau'(\tau), \quad x^i \rightarrow {x^i}' (\tau, x),
\end{align}
that comprises time-reparametrizations, and time-dependent spatial diffeomorphisms. The preferred time aligned with the aether flow is given in this case by
\begin{align}\label{eq:preferred_time}
    \tau = t+\int \frac{U_r}{U_t} {\rm d}r.
\end{align}
Note that $\tau'(\tau)$ must be a monotonous function to meet the chronology condition. We can similarly introduce a preferred spatial coordinate that diagonalizes the spatial metric by
\begin{align}\label{eq:gauge_cond}
    \rho = t + \int \frac{S_r}{S_t}{\rm d}r,
\end{align}
where the space-like vector $S^\mu$ is by construction orthogonal to the aether and reads
\begin{align}
    S_\m {\rm d}x^\m=-\frac{1-F(r)A(r)^2}{2A(r)}{\rm d}t-\frac{B(r)(1+F(r)A(r)^2)}{2A(r)F(r)}{\rm d}r.
\end{align}

As a direct consequence of working in the preferred coordinate neighborhood $(\tau,\rho,\vartheta,\varphi)$, the line-element naturally assumes the ADM form 
\begin{align}\label{eq:ADM_metric}
    {\rm d}s^2=(N^2-N_iN^i){\rm d}\tau^2 -2 N_i {\rm d}x^i {\rm d}\tau - \gamma_{ij}{\rm d}x^i {\rm d}x^j,
\end{align}
where $N,N^i$ and $\gamma_{ij}$ are the lapse, shift vector, and the metric induced in $\Sigma$, respectively. In the gauge \eqref{eq:gauge_cond} they read
\begin{align}
   N=\frac{1+F(r)A(r)^2}{2A(r)},\quad  N^i=0, \quad \gamma_{ij}{\rm d}x^i {\rm d}x^j=\left(\frac{1-F(r)A(r)^2}{2A(r)}\right)^2 {\rm d}\rho^2 +r^2(\tau,\rho) {\rm d}\mathbb{S}_2.
\end{align}

Note that the causal structure of this space-time can be viewed as an extremal limit of its general relativistic counterpart. Since violations of Lorentz invariance may permit superluminal propagation, i.e.~they allow for causal curves that are considered to be space-like in terms of GR, the Killing horizon becomes permeable in both ways. Effectively, the causal cone is widened up to the extent that the causal structure becomes Newtonian, i.e.~the causal cone's boundary aligns with $\Sigma$. However, it should be mentioned that besides this difference, the asymptotic regions coincide with the general relativistic ones \cite{Bhattacharyya16}. 

Remarkably, the anatomy of the black hole interior admits an additional horizon, called universal horizon (UH), at position $r=r_{\rm UH}$ determined by
\begin{align}\label{eq:UH_def}
    N=U\cdot \chi=0,\quad A\cdot \chi\neq 0,
\end{align}
where $A^\m=U^\n \nabla_\n U^\m$ is the acceleration of the aether \cite{Bhattacharyya16}. Due to the time-like character of the aether, it is easy to see that $U \cdot \chi=0$ requires the Killing vector to be spacelike. Hence, the UH must lie always inside the KH\footnote{This applies to outer horizons, e.g. black or white hole horizons. More complicate configurations occur in the case of multiple horizons, see e.g.~\cite{Mazza23}}. The conditions \eqref{eq:UH_def} define a constant radius surface with constant non-vanishing surface gravity (non-degenerate horizon) given by
\begin{align}\label{eq:kappa_UH}
    \kappa_{\rm UH}=-\frac{1}{2}(A\cdot \chi) \biggr|_{\rm UH}=-\frac{1}{2}U^\mu \partial_\mu (U\cdot \chi)\biggr|_{\rm UH} \,,    
\end{align}
and that does not reach spacelike infinity $i^0$. Being a compact surface, orthogonal to the preferred time-direction $U_\mu$, none of the curves which penetrate the UH can escape from it without crossing the same surface again, namely without going backwards in preferred time. For a causal (future-directed) trajectory, it is then impossible to reach $i^0$ from the region inside the UH, and thus the UH is a trapping horizon for every signal . This is supported by the explicit computation of the expansion coefficient $\theta_S$ along $S^\m$, which gives a characterization of the UH via $\theta_S=0$ -- see \cite{Carballo_Rubio_2022}.

\begin{figure}
	\includegraphics[scale=0.55]{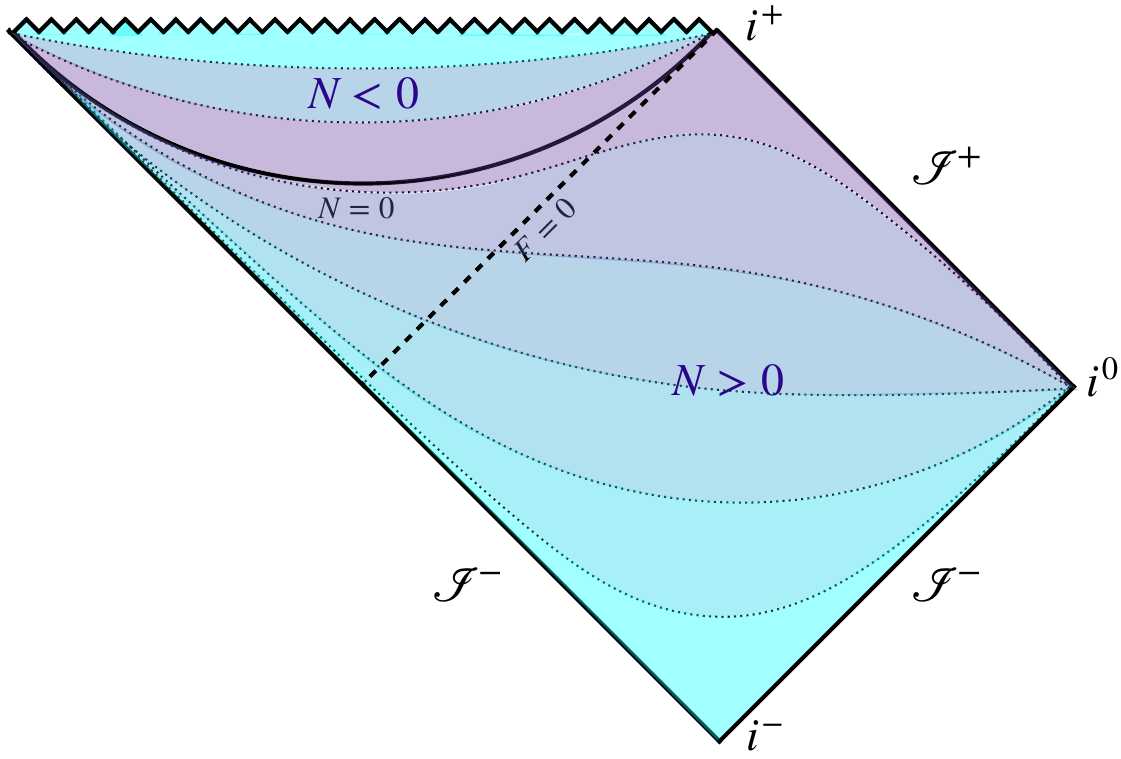}
	\caption{\label{PenAeS} 
Penrose diagram for an EA spherical black hole: the Killing horizon is depicted by the dashed line, while the bold line displays the universal horizon at $N=0$. Dotted lines mark the individual leafs $\Sigma\perp U$, with the color coding indicating the value of the preferred time from an outside observer perspective. Purple indicates large $\tau$, while cyan corresponds to small values of $\tau$. The asymptotic regions can be defined similarly to the relativistic Schwarzschild black hole \cite{Bhattacharyya16}.}
\end{figure}

Figure \ref{PenAeS} shows the corresponding Penrose diagram of this geometry. The character of the foliation changes at exactly the set of points that defines the universal horizon. For $r\le r_{\rm UH}$, the leafs reach $i^+$, instead of $i^0$, meaning that infinite speed signals straddle those hypersurfaces forever without ever being able to escape the enclosure. Hence, the universal horizon has compact topology homeomorphic to $\mathbb S_2$, and marks a surface of simultaneity because of the aether's hypersurface orthogonality. When the time coordinate is chosen to be \eqref{eq:preferred_time}, the UH is mapped to $\tau\rightarrow \infty$. This implies that the foliation describing the exterior of the black hole cannot be globally extended into its interior. Instead, a second foliation covers the region $r<r_{\rm UH}$, with its globally defined time also diverging at the position of the UH, but acquiring a finite value at the singularity \cite{DelPorro22}.

Note that a constant nonzero surface gravity \eqref{eq:kappa_UH} implies that $\partial_r N(r_{\rm UH})\neq 0$, so that the lapse vanishes linearly at the UH. Hence, $N$ changes sign in the interior of the UH, which leads outward observers to conclude that the preferred time runs backwards within that region \cite{DelPorro22}. 

In what follows, and in order to perform calculations that can be easily compared with the general relativistic situation, we particularize our investigation for the functions $F(r)$, $B(r)$, and $A(r)$ to
\begin{align}\label{eq:solution_functions}
    F(r)=1-\frac{2M}{r}, \quad B(r)=1, \quad A(r)=1,
\end{align}
so that the metric corresponds to the standard Schwarzschild metric. This enables direct comparisons with General Relativity in a simple manner. Note however that this metric corresponds to a particular corner in the parameter space of EA gravity -- see \cite{Barausse:2011pu} -- which is however excluded by the most updated observational constraints \cite{Gupta:2021vdj}. Thus, this solution cannot represent a realistic one. Nonetheless its simplicity allows for a test of our results in a transparent way. In spite of that, we remark that, although operationally more involved, all the conclusions attained in the following can be extrapolated to any arbitrary solution of the form \eqref{eq:metric} with aether \eqref{eq:aether}, including those in phenomenologically viable corners of the parameter space.

\subsection{Lorentz-violating matter}\label{s:LVmatter}

We will probe the black hole geometry via a scalar field $\phi$ coupling to both the metric and the aether, and endowed with an anisotropic scaling of the form \eqref{eq:Lifshitz_scaling}. This corresponds to a Lifshitz scalar field, whose action in curved-space reads
\begin{align}\label{eq:lifshitz_action}
    S_{\rm m}=-\frac12\int {\rm d}^4x \sqrt{-g}\left[\partial^\m\phi\partial_\m\phi+\sum_{j=2}^n\frac{\beta_{2j}}{\Lambda^{2j-2}}\phi \Delta^j\phi\right],
\end{align}
and leads to the equation of motion
\begin{align}\label{eq:lifhistz_eom}
    \square \phi -\sum_{j=2}^n\frac{\beta_{2j}}{\Lambda^{2j-2}}\Delta^j\phi=0.
\end{align}
Here all the $\beta_i$ are coupling constants, while $n$ controls the scaling of the equations at large momentum. We have normalized $\beta_{2n}=1$, which is always possible by rescaling the energy scale $\Lambda$, and also set the IR speed of the mode to $c=1$, in order to agree with the GR result in the decoupling limit. We will also assume that all $\beta_i \ge 0$, in order to avoid regions of the parameter space with sub-luminal behavior. The d'Alembertian and the induced Laplace operator are given by $\square=g^{\m\n}\nabla_\m\nabla_\n$ and $\Delta=\gamma^{\m\n}\nabla_\m\nabla_\n$, with $\gamma^{\m\n}=g^{\m\n}-U^\m U^\n$ the projector onto the spatial leafs.

The higher spatial derivative terms in \eqref{eq:lifhistz_eom} endow the scalar field with a modified dispersion relation at high energies, while avoiding the presence of ghosts. This can be easily confirmed by looking at a spherical wave around flat space. Inserting a flat metric in spherical coordinates, a trivial aether field $U^\m=(1,\Vec{0})$, and working in Fourier space -- so that the s-wave is $\phi\propto\exp\left(-i \omega t+ i k r\right)$ -- the equation of motion \eqref{eq:lifhistz_eom} leads to a modified dispersion relation
\begin{align}\label{eq:disp_flat}
    \omega^2=k^2+\sum_{j=2}^n\frac{\beta_{2j}}{\Lambda^{2j-2}}k^{2j},
\end{align}
which is indeed Lorentz-violating for scales $k\gtrsim \Lambda$, but only contains two causal solutions $\omega(k)$. This is of course a consequence of the fact that the aether field is aligned with the time direction, which reduces the induced Laplace operator $\Delta$ to a purely spatial one. Coherently with what we have discussed in the previous section, performing an FDiff transformation \eqref{eq:FDiff} will not change the shape of the dispersion relation, while a generic Diff transformation would map \eqref{eq:disp_flat} into a polynomial of order $2n$ in both $\omega$ and $k$.


\section{Matter kinematics}
\label{sec:MattKin}

Hereinafter we focus on solutions to \eqref{eq:lifhistz_eom} in the WKB regime, namely
\begin{align}\label{eq:wkb_ansatz}
    \phi_{\rm WKB}= \phi_0 e^{i {\cal S}_0} \,,
\end{align}
where the phase ${\cal S}_0 $ corresponds to the lowest-order contribution in the $\hbar$ expansion of the solution (see \cite{Kerner08,banerjee2008quantum} and references therein). Within the geometrical optics approximation, ${\cal S}_0 $ can also be interpreted as a point particle action, so that curves with constant phase are equivalent to trajectories followed by classical rays. Derivatives of ${\cal S}_0$ can thus be described through the four-momentum $p_\mu$ of a particle
\begin{align}\label{eq:classical_action}
     {\cal S}_0=- \int p_\mu {\rm d}x^\mu= -\int \omega \, U_\mu {\rm d}x^\mu - \int k \, S_\mu {\rm d}x^\mu \,,
\end{align}
which we have decomposed in the preferred frame chart parameterized by $U_\m$ and $S_\m$. The functions $\omega$ and $k$ thus take the role of energy and radial momentum within this frame. They satisfy the eigenvalue equations $U^\mu \partial_\mu \phi=-i \omega \phi$ and $S^\mu \partial_\mu \phi= i k \phi$. Note that we have suppressed any angular dependence, thus focusing solely on the s-wave contribution to the solution.

The advantage of this choice is that, as previously discussed, only in the preferred frame the equation of motion reduces to the dispersion relation
\begin{align}\label{eq:phi_eom}
     \omega^2 = k^2  + \sum_{j=2}^{n}\frac{\beta_{2j}}{\Lambda^{2j-2}} k^{2j} + G(\omega, \nabla \omega,k, \nabla k)\,,
\end{align}
where the function $G(\omega, \nabla \omega,k, \nabla k)$ encodes all the terms which depend on the derivatives of either $\omega$ or $k$. From now on we will also assume an adiabatic evolution of the field, so that $|\nabla k| \ll k^2$. This allows to neglect the contribution from $G(\omega, \nabla \omega,k, \nabla k)$. Of course, adiabaticity will not hold everywhere, but we will discuss its range of validity later. For now, let us assume that it is a valid approximation within the whole radial domain, so that the dispersion relation simplifies to the flat space-time version \eqref{eq:disp_flat}.

Importantly, note that $\omega$ and $k$ are space-time dependent for a given solution, since neither $U^\m$ nor $S^\m$ are Killing vectors. Fortunately, the presence of the time-like Killing vector $\chi^\m \partial_\m = \partial_t$ allows to separate the variables in the $(t,r)$ plane via the eigenvalue problem $\chi^\m \partial_\m \phi_{\Omega}=-i \Omega \phi_{\Omega}$, where the Killing energy $\Omega$ is now constant for each mode $\phi_\Omega$. Rewriting the left hand side of this definition in the preferred frame, we arrive at the identity
\begin{align}\label{eq:KE}
     \Omega = \omega (U \cdot \chi) + k (S \cdot \chi)= \omega N + kV\,,
\end{align}
where we have introduced the tangential flow of the foliation $V=(S\cdot\chi)=-M/r$. Together, equations \eqref{eq:disp_flat} and \eqref{eq:KE} provide a complete system that can be solved for $\omega$ and $k$, and hence for the monochromatic mode $\phi_\Omega$.

\subsection{Mode analysis}
Solving \eqref{eq:KE} for $\omega$ and substituting it into the dispersion relation \eqref{eq:disp_flat}, we obtain the following algebraic equation for $k$
\begin{align}\label{eq:krho_eq}
     \sum_{j=2}^{n}\frac{\beta_{2j}}{\Lambda^{2j-2}} k^{2j} + \biggr( 1- \frac{V^2}{N^2} \biggr) k^2 + \frac{2 \Omega V}{N^2} k - \frac{\Omega^2}{N^2} =0\,.
\end{align}
This is a polynomial of degree $2n$ with position dependent coefficients. Note however that all coefficients accompanying powers of $k$ are sign definite for all $r$ -- once the sign of $\Omega$ is chosen --, except for the term multiplying $k^2$, which vanishes at the Killing horizon, since $|\chi|^2=N^2-V^2$. Using Descartes's rule of signs, this implies that in the outer region of the Killing horizon, \eqref{eq:krho_eq} will display exactly two real roots, corresponding to positive and negative values of $k$. Indeed, in the asymptotic region $r\rightarrow \infty$, where $N=1$ and $V=0$, we get
\begin{align}\label{eq:asympt_eom}
     \sum_{j=2}^{n}\frac{\beta_{2j}}{\Lambda^{2j-2}} k^{2j} +  k^2 - \Omega^2=0\,,
\end{align}
which has exactly two real roots of opposite sign, that can be obtained perturbatively in $\Lambda$. 

In the region inside the KH instead, the change of sign allows for the number of real roots to potentially grow up to four. Focusing in the near-UH limit, where $N \to 0$, we indeed find four roots \cite{Parentani15,Herrero-Valea20} -- two \textit{hard solutions}, for which $k$ diverges when approaching the UH; and two \textit{soft solutions}, for which $k$ remains finite. The latter two can be obtained straightforwardly by setting $N=0$ and $V=-1$ in \eqref{eq:KE}, obtaining 
\begin{align}\label{eq:softsol}
    k_{s}=-\Omega,\quad  \omega_{s}=\pm \Omega \sqrt{1 + \sum_{j=2}^n \beta_{2j}\alpha^{2j-2}},
\end{align}
where we have introduced the dimensionless quantity $\alpha= \Omega/\Lambda$, that will be useful in the remainder.

For hard solutions instead, the radial momentum $k$ diverges at the position of the horizon, with its behavior controlled by the highest power in the dispersion relation. Its exact shape can be obtained perturbatively by using boundary layer theory \cite{bender1999advanced}. Hence, we rescale the momentum as $k=\Theta(N) \hat k$ and look for values of $\Theta(N)$ that balance different terms in the equation, solving afterwards for $\hat k$. This method unveils four non-vanishing solutions. Two of them correspond to the previous soft modes, while the other behave as
\begin{align}\label{eq:hardsol_out}
     k_{h,{\rm out}}^{\pm}= \pm \Lambda  \biggl(-\frac{ V}{N}\biggr)^{ \gamma} - \frac{1 }{n-1} \frac{\Omega}{V} + O(N)\,, \qquad \omega^{\pm}_{h,{\rm out}}= \pm \Lambda \biggl(-\frac{ V}{N}\biggr)^{n \gamma} + \frac{n }{n-1} \frac{\Omega}{N} + O(N^0)\,,
\end{align}
where we have introduced the exponent $\gamma=(n-1)^{-1}$. The out label indicates that these solutions are obtained in the limit $N\rightarrow 0^+$, corresponding to the outer near-horizon region. If we look at the inner neighborhood instead, we get
\begin{align}\label{eq:hardsol_in}
     k_{h,{\rm in}}^{\pm}= \pm \Lambda  \biggl(\frac{ V}{N}\biggr)^{ \gamma} - \frac{1 }{n-1} \frac{\Omega}{V} + O(N)\,, \qquad \omega_{h,{\rm in}}^{\pm}= \mp \Lambda \biggl(\frac{ V}{N}\biggr)^{n \gamma} + \frac{n }{n-1} \frac{\Omega}{N} + O(N^0)\,.
\end{align}

At some intermediate point inside the Killing horizon, but not necessarily at its surface, the character of the equation must change and the number of real solutions reduces to two, connecting to the exterior modes. However, the position of this turning point depends non-trivially on the Killing energy $\Omega$, as we shall see.

\subsection{Following a wave-packet}

While above we have discussed monochromatic waves, in what follows we focus on the more physical case of a wavepacket. This has a direct link with the modelling of a particle and, as we will see in the following, it allows us to explore the differences induced by the modified dispersion relation \eqref{eq:phi_eom} with respect to the general relativistic case.

From now on, we model the radiation climbing the gravitational well as a Gaussian profile built from a set of positive energy modes $\phi_\Omega$, centered around a preferred frequency $\omega_0$
\begin{align}\label{eq:wavepacket}
    \psi_{\omega_0}(r) = \int_0^{+ \infty} \frac{{\rm d} \omega}{\sqrt{2 \pi \sigma}} \, \phi_\Omega(r) \, e^{-\frac{(\omega-\omega_0)^2}{2 \sigma}} = \int_0^{+ \infty} \frac{{\rm d} \Omega}{\sqrt{2 \pi \sigma}}  \frac{{\rm d} \omega }{{\rm d} \Omega} \, \phi_\Omega(r) \, e^{-\frac{(\omega-\omega_0)^2}{2 \sigma}} ,
\end{align}
with standard deviation $\sigma$, and where ${\rm d} \omega/{\rm d}\Omega$ can be computed from the behavior of the modes. This expression satisfies the equations of motion \eqref{eq:phi_eom} as long as the approximations previously introduced -- WKB and adiabaticity -- hold, since mode-mode interaction is negligible in that case. Note that at large radii $\omega\rightarrow \Omega$; this describes a Gaussian wave-packet in Killing frequency.

The wave-packet $\psi_{\omega_0}$ can be classically thought of as an object travelling with speed determined by its enveloping wave-front, with group velocity $ c_g(r,\a)= {\rm d} \omega/{\rm d} k$. This determines the four-velocity of the packet 
\begin{align}\label{eq:tangvec}
  \hat{v}^\m=U^\m + c_g S^\m  \,,
\end{align}
as well as its causal cone, after inserting one of the four explicit solutions $k(r)$.

\begin{figure}
	\includegraphics[scale=0.5]{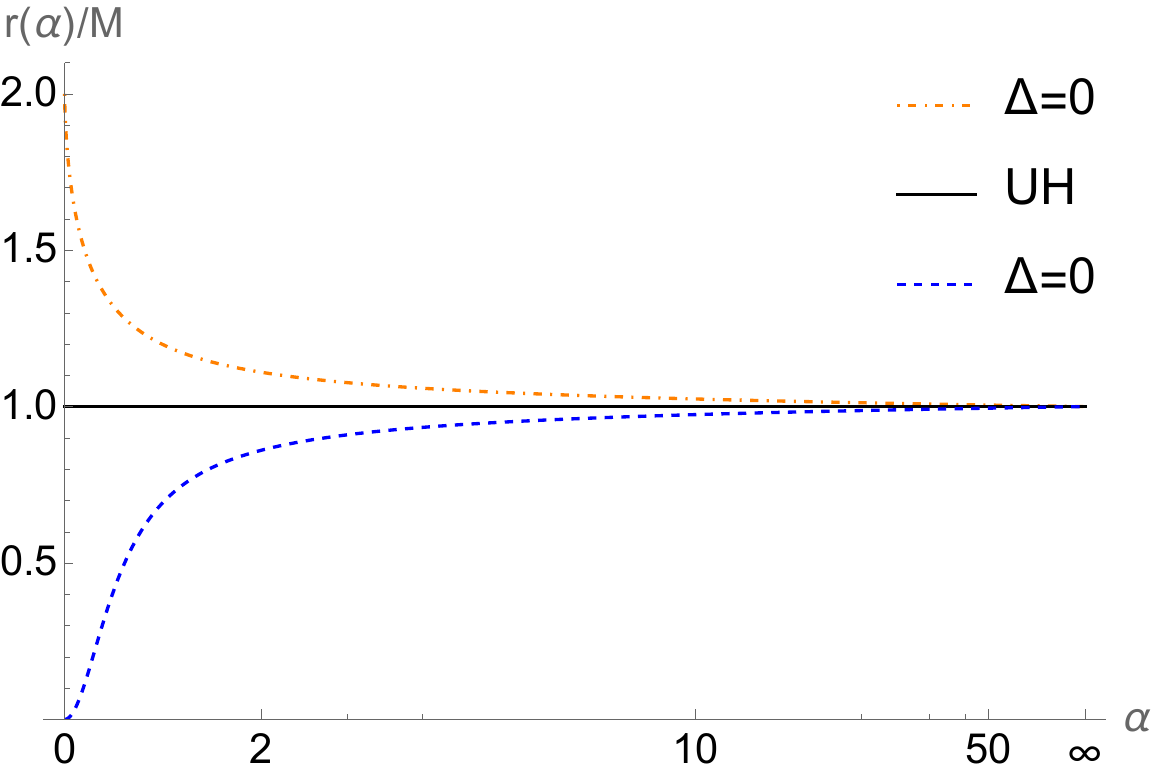}
	\caption{\label{f:turningpoint} Position of the roots $\hat{r}_{\rm out}(\a)$ (in orange, dash-dotted) and $\hat{r}_{\rm in}(\a)$ (in blue, dashed) in terms of $\a>0$. The solid black line represent the UH radius.}
\end{figure}

In order to get an analytical and quantitative grasp of the behavior of $\psi_{\omega_0}$ in all regions of space-time, we restrict ourselves to the case $n=2$ in \eqref{eq:lifshitz_action}, so that 
\begin{align}\label{eq:cg}
  c_g(r,\a)= \frac{k + 2 k^3/\Lambda^2}{\sqrt{k^2 + k^4/\Lambda^2}}=\frac{k}{\omega}\biggl(1+2\frac{k^2}{\Lambda^2} \biggr).
\end{align}
This will not spoil the generality of our conclusions, but it will allow us to perform explicit computations in a more transparent way. Also, the case $n=2$ corresponds to the lowest order in the effective field theory expansion for a CPT-even Lorentz violating theory \cite{Liberati13}. Since our results here will be perturbative, it suffices to retain this contribution without any loss of generality. For this choice, the equations of motion of the scalar field are fourth order in the radial momentum $k$, and the dispersion relation \eqref{eq:krho_eq} hence becomes a fourth order polynomial, which can be solved exactly if needed -- although its solutions are generically quite involved.

As we discussed in the previous section, the dispersion relation \eqref{eq:krho_eq} changes character in between the universal and Killing horizons, with two of its roots becoming complex. For $n=2$, this can be seen explicitly by looking at the discriminant of the corresponding polynomial\footnote{The discriminant $\Delta$ of a polynomial $\mathcal{P}(x,b_i)$ with real coefficients $\{ b_i \}$ is a real function of the coefficients $\Delta[\mathcal{P}](b_i)$. Its vanishing indicates degeneracy of at least two of the solutions, while a change of sign in $\Delta$ implies that two real solutions become complex conjugate.} in \eqref{eq:krho_eq}
\begin{align}
\label{eq:deltaroot}
  \Delta(r, \a)=\left(4 \alpha ^2+1\right)^2 \hat{r}^4-2 \left(16 \alpha ^4+16 \alpha ^2+3\right) \hat{r}^3+4 \left(4 \alpha ^4+17 \alpha ^2+3\right) \hat{r}^2-8 \left(9 \alpha ^2 +1 \right) \hat{r}+27 \alpha ^2\,,
\end{align}
where $\hat{r}=r/M$. Remarkably, $\Delta(r,\a)$ displays two real roots $\{\hat{r}_{\rm out}(\a),\hat{r}_{\rm in}(\a)\}$, one on each side of the UH, and whose positions depend on $\alpha$, as it can be seen in Fig.\ref{f:turningpoint}. Note that for $\alpha \to 0$, the outer solution approaches the Killing horizon, moving inwards as we increase the value of $\alpha$. The inner one, instead, starts from the singularity and moves towards the UH for large $\alpha$. 

\begin{figure}
	\centering
	\includegraphics[scale=0.27]{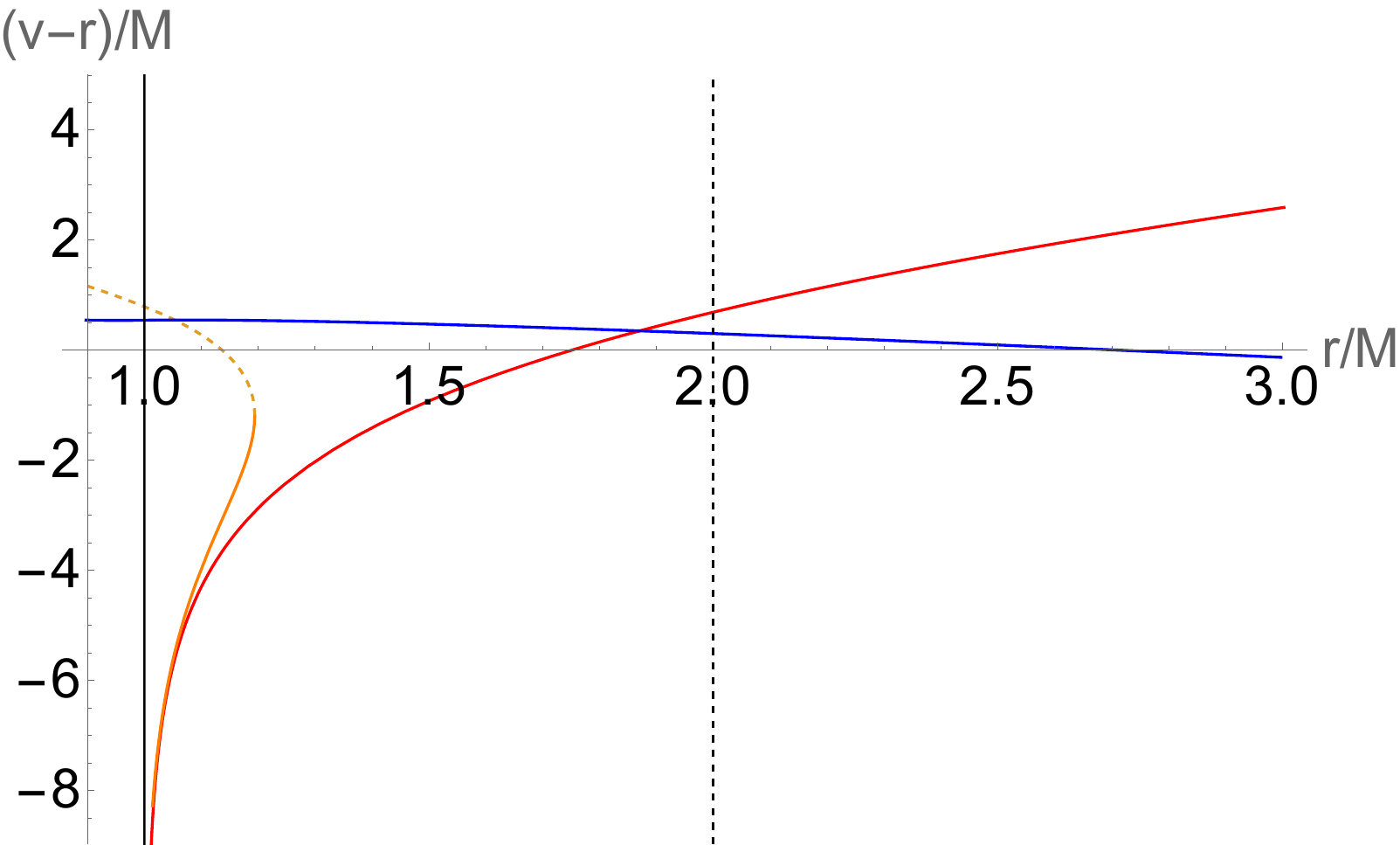}
	\includegraphics[scale=0.27]{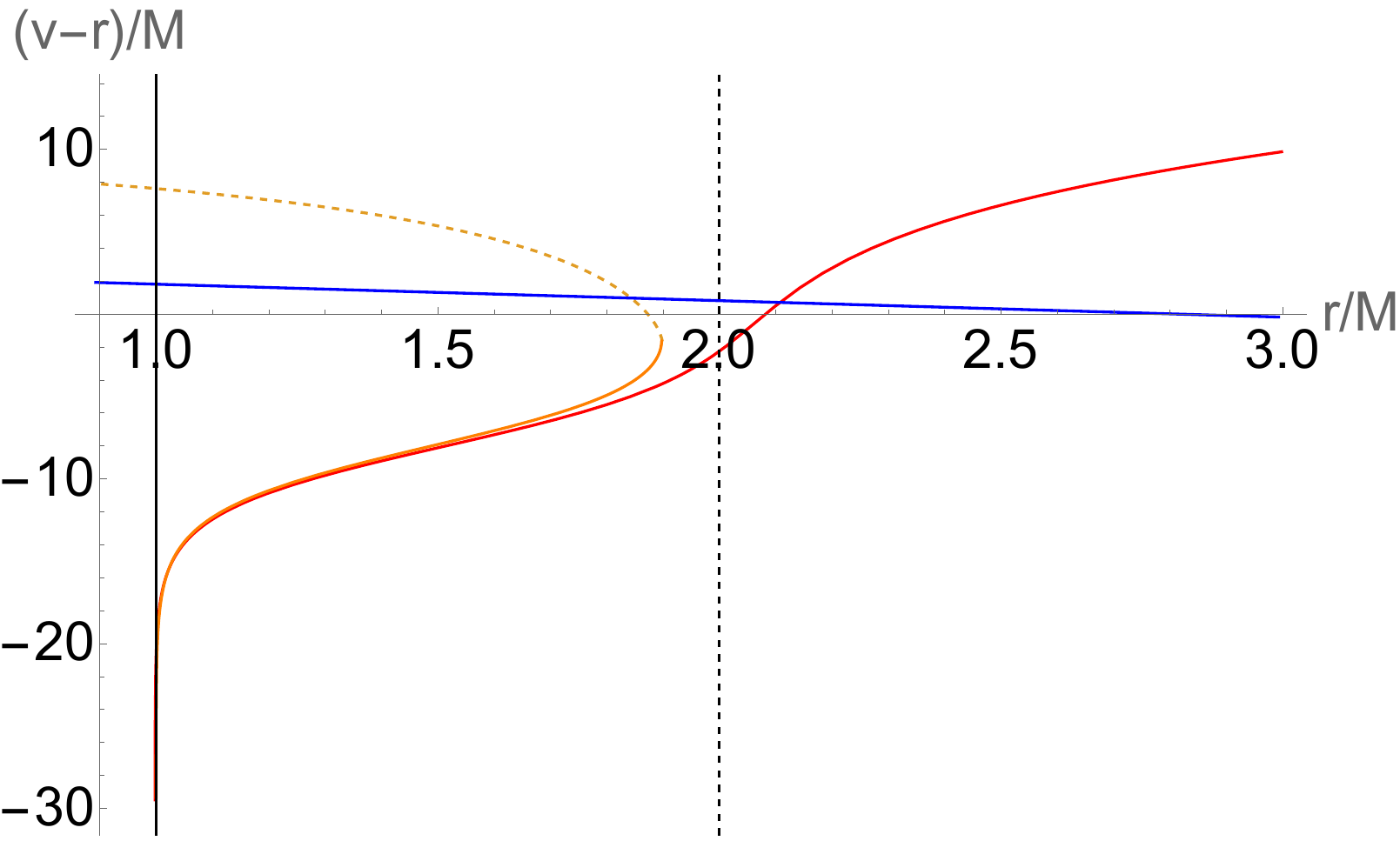}
	\caption{Characteristics of the four solutions (blue, red, solid orange and dashed orange lines) for $r>r_{\rm UH}$, corresponding to the integral lines of (\ref{eq:tangvec}) evaluated with the four different shapes of $c_g$, in the $(t^*,r)$ plane, where $t^*=v-r=t-2M \log(|1-2M/r|)$, and $v$ is the usual Eddington-Finkelstein coordinate. The left plot is evaluated at $\alpha=1$, while the right one uses $\alpha=10^{-2}$. The KH is at $r/M=2$ (dashed black vertical line), while the UH (solid black vertical line) sits $r/M=1$.}
	\label{f:characteristics}
\end{figure}

The degeneracy of solutions is directly reflected in the propagation of the rays. Indeed, once the vectors $U_\m$ and $S_\m$ are given, the trajectory of the wave-packet is completely determined by the group velocity \eqref{eq:cg}, and a degeneracy in $k$ logically implies a degeneracy in the trajectory. This means that the radius $r_{\rm out}(\a)$ actually corresponds to the turning point where two solutions meet, as depicted in Fig.\ref{f:characteristics} for the dashed and solid orange lines. Both solutions can be considered as a single continuous trajectory which peels out from the UH but remains trapped within the inner region, eventually falling back into the UH \cite{Parentani15}. The other two modes -- red and blue -- have support everywhere for $r>r_{\rm UH}$. The latter represents an in-going mode, while the former corresponds to an out-going trajectory, peeling off the UH and climbing up to $\mathscr{I}^+$. If pure out-going boundary conditions are set, this is the only mode allowed in the asymptotic region. Thus, it will describe the characteristics of the wavepacket \eqref{eq:wavepacket} which we want to trace, corresponding to radiation emitted in the neighborhood of the UH and climbing the gravitational well.

\subsection{Analysis of the solutions}

\begin{figure}
	\includegraphics[scale=0.25]{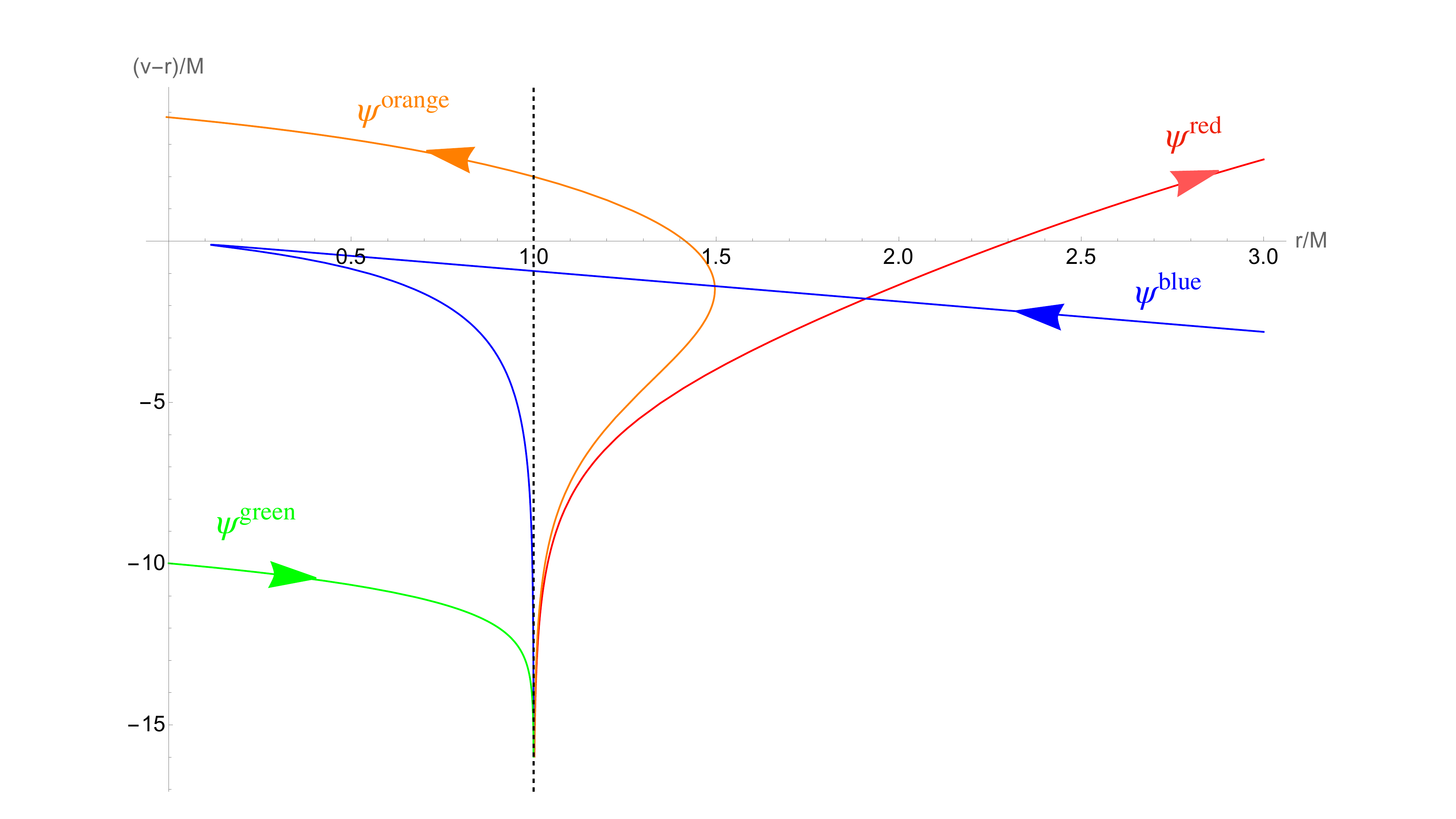}
	\caption{\label{f:allmodes} Characteristics of all modes for $\alpha=0.2$. Starting from outside the Killing horizon at $r/M=2$, we see that at fixed $r$, there are only two possible solution: the red out-going mode and the blue in-going one. In the interior of the Killing horizon we find the turning point $r_{\rm out}(\a)$, where the two orange lines meet. For $r_{\rm UH}<r<r_{\rm out}(\a)$ we have four real solutions. Inside the UH, the red line and the hard branch of the orange mode leave the stage to the green mode and the hard branch of the blue mode, until they reach $r_{\rm in}(\a)$, where the two blue modes' branches are linked. Beyond this point, only the orange and green lines hit the singularity. The arrows represent the direction of propagation of the rays, given by their group velocities $c_g$. The labels $\{\psi^{\rm red},\psi^{\rm orange},\psi^{\rm blue},\psi^{\rm green} \}$ associated to each ray are the same one that we refer to along the text.}
\end{figure}

Let us now focus on discussing the properties of the four different solutions for $c_g$, obtained by plugging the soft and hard solutions \eqref{eq:softsol} and \eqref{eq:hardsol_out} into \eqref{eq:cg}. Before going into detail, let us recall that the discriminant \eqref{eq:deltaroot} vanishes at two points, one outside and one inside the UH. This means that the qualitative behavior of the modes will be similar in both regions. Indeed, this is clearly seen in Fig.\ref{f:allmodes} (notice the similarity of this figure with the one in \cite{Parentani15}), where the four modes are plotted for $r>0$. As we can see, the inner region of the UH also displays a mode that peels next to the horizon, one mode that crosses, and two modes that combine to a smooth curve at the turning point. 

Although qualitatively the same, the details of the solutions in both regions are different. In particular, the energy carried by the different modes, and their direction of propagation, will depend on the side of the horizon where they have support, as described by their four-velocity. Using the decomposition of the latter in \eqref{eq:tangvec}, we can obtain the equation of motion for the trajectory as simply
\begin{align}
\label{eq:trajectories}
   \frac{ \hat{v}^\rho}{\hat{v}^\tau}= \dot{\rho}= - \frac{N}{V} c_g(r, \a) \,,
\end{align}
where $\dot{\rho}={\rm d}\rho/{\rm d}\tau$. Plugging $c_g$ onto this expression allows to obtain the curve $(\tau,\rho(\tau))$ describing the trajectory of the corresponding wave-packet in preferred frame coordinates. Furthermore, we define a local effective null coordinate 
\begin{align}\label{eq:baru}
    {\rm d}\bar u=\dot \rho {\rm d}\tau-{\rm d}\rho,
\end{align}
so that field trajectories correspond to lines of constant $\bar u$. This can be expressed as
\begin{align}\label{eq:trajectories2}
    0={\rm d} \bar u =( \Dot{\rho}-1) {\rm d}v + \biggl( \Dot{\rho} \frac{U_r}{U_v} - \frac{S_r}{S_v} \biggr) {\rm d}r,
\end{align}
where we have introduced the Eddington-Finkelstein coordinate ${\rm d}v={\rm d}t + r/(r-2M) {\rm d}r$. Using \eqref{eq:trajectories} and evaluating the vectors $U^\m$ and $S^\m$ explicitly in this coordinate chart, the previous expression simplifies to
\begin{align}\label{eq:trajectory_vr}
    0={\rm d}v - \frac{c_g+1}{N c_g + V}{\rm d}r,
\end{align}
which provides the trajectory in $\{v,r\}$ coordinates.

\begin{figure}
	\includegraphics[scale=0.45]{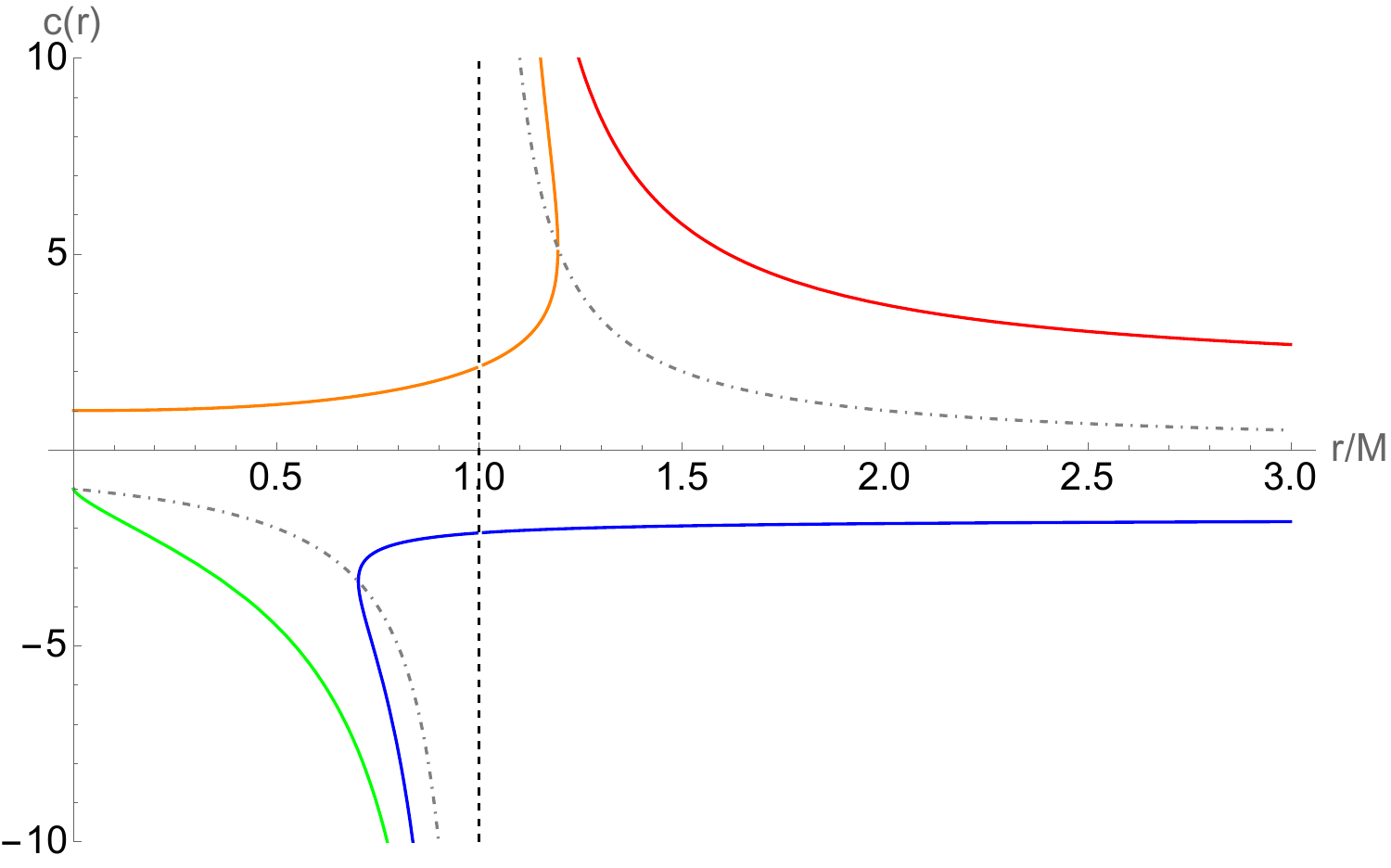}
	\caption{\label{f:groupvel} Group velocities $c_g(r,\alpha)$ for $|\alpha|=1$. The red and blue modes are computed at $\alpha=1$, while the orange and green ones have $\alpha=-1$, corresponding to negative Killing energy (cf. below). The gray dot-dashed line corresponds to the function $-V/N$. Its intersections with the group velocities mark the points where $\Dot{\rho}=1$. Those happen exactly at the turning points for the blue and orange modes.}
\end{figure}

It is easy to check that the second term in \eqref{eq:trajectories2} has definite sign for $|c_g|\geq 1$. Hence, the direction of propagation of the rays is controlled only by the first term, which depends upon $\dot \rho$. Whenever $\dot \rho$ crosses unity, the mode reverses its trajectory. This indicates that $\dot \rho=1$ corresponds to the turning point in the evolution of the rays. From \eqref{eq:trajectories} we see that this happens only when $c_g(r,\a)=-V/N$ and, as we can observe in Fig.\ref{f:groupvel}, only the blue and orange modes fulfill this property. Note however that the sign of $\omega$ and $k$ -- see e.g. \eqref{eq:hardsol_out} and \eqref{eq:hardsol_in} -- depends on the side of the UH where they are evaluated, since $N$ vanishes and changes sign at that point. 

\subsubsection{Outside the UH} 
This region corresponds to $N>0$, and includes the asymptotic region $r\rightarrow \infty$, where only two real solutions exist for $\Omega>0$, both with $\omega=\Omega=\sqrt{k^2+k^4/\Lambda^2}$, and $\pm k$. From \eqref{eq:cg} we see that they have opposite group velocity and hence opposite direction of propagation. They correspond to the red and blue modes $c_g^{\rm red}>0$, and $c_g^{\rm blue}<0$. As a consequence, when increasing $\tau$ in the preferred frame of coordinates, the blue mode moves along decreasing values of $\rho$, while the red mode travels in opposite direction.
 
Both modes can be smoothly continued towards the interior of the Killing horizon, matching with the two real roots of \eqref{eq:krho_eq} always present for $r>r_{\rm UH}$. Once at the UH, as previously discussed, the red ray peels out, while the blue one crosses it. Since both modes correspond to $\Omega>0$, it is immediate to recognize that the blue ray carries an energy $\Omega$ from past infinity into the UH, while the red one extracts an equivalent amount from it, eventually reaching the asymptotic observer at future infinity.

Inside the Killing horizon we find the turning point $r_{\rm out}(\a)$, so that another two real solutions are possible for $r_{\rm UH}<r< r_{\rm out}(\alpha)$. Those correspond to the orange mode $\psi^{\rm orange}$, which is split into two branches. One of them is hard, corresponding to the partner of the red mode in \eqref{eq:hardsol_out}, while the other one is soft. In both cases, the orange mode corresponds to $\omega<0$. However, by definition, $\omega$ respects the identity
\begin{align}\label{eq:schroedinger_omega}
    \partial_\tau \phi=-i N \omega \phi,
\end{align}
which requires $N\omega>0$ for the particle to evolve forward in the preferred time $\tau$. In this region of space-time, this translates to $\omega>0$ and thus poses a problem for the orange mode. A way out of this is to employ the CPT symmetry of the field equations, given by the inversion of the triplet $(\Omega, \omega,k) \to (-\Omega,-\omega,-k)$. In doing so, we obtain a CPT reversed mode with the same group velocity $c_g^{\rm orange}$, but with positive, preferred energy $\omega >0$ while $\Omega<0$\footnote{In \cite{Parentani15}, this property is taken into account through the Hermitian conjugation of the modes.}. 

Using \eqref{eq:hardsol_out}, \eqref{eq:cg}, and  \eqref{eq:trajectories}, we can observe that the hard branch has $\dot \rho>1$, therefore moving outwards, until it reaches the turning point $r_{\rm out}(\alpha)$, where $\Dot{\rho}=1$ and it merges with the soft branch, which extends the mode through the region where $\dot \rho<1$ -- see Fig.\ref{f:groupvel}. The whole path corresponds to an evanescent process where the mode extracts a quantum of energy $\Omega<0$ from the horizon, but eventually falls down back onto the black hole, restoring its energy. Indeed, since it has negative Killing energy, the orange mode is restricted to live in the region behind the Killing horizon, since negative $\Omega$ cannot be on-shell on its exterior. A full summary of the behavior of all modes can be found in Table \ref{t:modes_out}.

\begin{table}[h!]
    \centering
    \begin{tabular}{|c|c|c|c|}
    \hline
    &  sgn($\Omega$)  & sgn($\omega$)  & Direction\\
       \hline
    $\psi^{\rm red}$ & $+$ & $+$ & Out \\
     \hline
   $\psi^{\rm blue}$ & $+$ & $+$ &  In \\
    \hline
    $\psi^{\rm orange}_{\rm hard}$ & $-$ & $+$ &  Out \\
     \hline
     $\psi^{\rm orange}_{\rm soft}$ & $-$ & $+$ &  In \\
\hline
    \end{tabular}
    \caption{Behavior of the modes in the exterior region. The last column indicates the direction of propagation with respect to the gravitational well.}
    \label{t:modes_out}
\end{table}

\subsubsection{Inside the UH} 

Inside the UH, the discussion unfolds in a similar way to the exterior region. The important difference is that the lapse changes sign \cite{DelPorro22}, i.e. $N<0$. This gives a different interpretation to (\ref{eq:schroedinger_omega}), which now implies that on-shell modes that evolve forward in the -- interior -- preferred time must have $\omega<0$.

Let us first extend the soft branches through the UH, carrying $\Omega>0$ for the blue mode, and $\Omega<0$ for the orange one. This fixes automatically the sign of $k$, since the conservation equation \eqref{eq:softsol} for the soft modes gives $k=-\Omega$. Moreover, in order to preserve the sign of $c_g$ \eqref{eq:cg} for both rays, we need to impose $\omega>0$, such that $N\omega<0$ is ensured. Relation \eqref{eq:schroedinger_omega} implies that those modes move backwards in the interior foliation's preferred time. Although it might seem counter-intuitive at first sight but this is actually coherent for a mode that crosses the UH from the exterior towards the interior, since interior and exterior preferred times are oriented in opposite directions \cite{DelPorro22}.

Going back to the modes, we understand from Fig.\ref{f:groupvel} and  \eqref{eq:trajectories}, that a turning point can only exist behind the UH for $c_g<0$, due to the change of sign of $N$. Hence, the orange mode will eventually hit the singularity with $\Omega<0$, while the blue mode will turn around at $r_{\rm in}(\alpha)$, connecting with a hard branch. This is selected from \eqref{eq:hardsol_in} by preserving $\Omega>0$, and thus corresponding to the solution with $\omega>0$ and $k<0$. Let us emphasize again, this mode is hence propagating backwards in the preferred time. It represents a ray carrying its positive Killing energy towards the horizon in the $(t^*,r)$ plane as it crosses leafs of decreasing lapse value.

Finally, there is an extra hard mode, represented in green, without turning point and with support only in the region $r<r_{\rm UH}$. It can be obtained from the exterior red mode by letting $(N,\Omega,k) \to (-N,-\Omega,-k)$. Considering this we obtain a ray with $\omega >0$ and $\Omega<0$ that propagates backwards in preferred time, towards the singularity. Again, a summary of the properties of all modes in this region can be found in Table \ref{t:modes_in}.

\begin{table}[h!]
    \centering
    \begin{tabular}{|c|c|c|c|}
    \hline
    &  sgn($\Omega$)  & sgn($\omega$) & Direction\\
       \hline
    $\psi^{\rm green}$ & $-$ & $+$ & In \\
     \hline
   $\psi^{\rm blue}_{\rm soft}$ & $+$ & $+$  & In \\
    \hline
    $\psi^{\rm blue}_{\rm hard}$ & $+$ & $+$  & Out \\
     \hline
     $\psi^{\rm orange}$ & $-$ & $+$  & In \\
\hline
    \end{tabular}
    \caption{Behavior of all modes in the interior region. The last column indicates the direction of propagation with respect to the UH.}
    \label{t:modes_in}
\end{table}

\section{Particle production}\label{s:PP}
After discussing the solutions to \eqref{eq:phi_eom} in all regions of space-time, we focus here on our initial goal of exploring the phenomenon of Hawking radiation within Lorentz violating gravity. To tackle this, we set purely out-going boundary conditions at $\mathscr{I}^+$, thus eliminating the blue mode from the spectrum, and we split the problem in two parts, regarding the production of radiation in the neighborhood of the UH, and its subsequent propagation to the asymptotic region, in the form of a wave-packet. Turning off all ingoing radiation has also the intuitive meaning of setting the asymptotic past infinity $\mathscr{I}^-$ to the Minkowski vacuum state, as we will see later on.

\subsection{Particle production at the Universal Horizon}

We thus focus here on the red mode, which is the only mode that has support in the whole region $r>r_{\rm UH}$ after imposing out-going boundary conditions. Close to the UH, this mode displays a peeling behavior, and is described in terms of hard modes only. Returning to arbitrary $n$ for the moment and recalling \eqref{eq:hardsol_out}, we have
\begin{align}
  \frac{{\rm d} \omega}{{\rm d} \Omega}= \frac{n}{n-1} \frac{1}{N} \,, \qquad \omega-\omega_0= \frac{n}{n-1} \frac{ \Omega - \Omega_0}{N}\,.
\end{align}

In this limit, the distribution in \eqref{eq:wavepacket} hence becomes a delta distribution, given the identity
\begin{align}
    \lim_{N \to 0^+} \frac{n}{n-1}\frac{1}{\sqrt{2 \pi \sigma}N}  \, e^{-\frac{n^2 (\Omega - \Omega_0)^2}{2 \sigma N^2 (n-1)^2}} = \delta(\Omega - \Omega_0) \,.
\end{align}
What we observe here, is nothing else than a gigantic blue-shift which erases the details of the wave-packet when traced back to the UH. In this limit, the wave-packet experiences infinite squeezing, such that it eventually degenerates to a monochromatic mode $\phi_{\Omega_0}$ near the UH, i.e.
\begin{align}
\label{eq:psi_squeezing}
    \lim_{r \to r_{\rm UH}^+} \psi^{\rm red}_{\omega_0}= \phi^{\rm red}_{\Omega_0}=\phi^{\rm red}_0 e^{i {\cal S}^{\rm red}_0},
\end{align}
with $\phi^{\rm red}_0$ being a constant.

Due to this feature, the motion of the wave-packet in the neighborhood of the UH is not anymore governed by its group velocity, but by the phase velocity $c_p$ of the monochromatic wave instead. Although this is also known to happen in the general relativistic case, in there group and phase velocities coincide due to relativistic invariance of the dispersion relation, which is not the case here.

Once the wave-packet is squeezed, its trajectory will be described by the curve $\gamma$ traced by the monochromatic wave, corresponding to constant phase contours of $\phi_{\Omega_0}$ labelled by an evolution parameter $\lambda$
\begin{align}
\label{eq:UH_traj}
    0= \frac{{\rm d} \phi_{\Omega_0}}{{\rm d} \lambda } \biggl|_\gamma= \frac{{\rm d} \tau}{{\rm d} \lambda } \biggl( \frac{{\rm d} \phi_{\Omega_0}}{{\rm d} \tau } +\frac{{\rm d} \rho}{{\rm d} \tau } \frac{{\rm d} \phi_{\Omega_0}}{{\rm d} \rho } \biggr) \biggl|_\gamma.
\end{align}
From here, we obtain the equation for the curve in the $\{\tau,\rho\}$ plane as simply
\begin{align}
\label{eq:monochromaric_causal_cone}
    \frac{{\rm d} \rho}{{\rm d} \tau }\biggl|_\gamma=- \frac{\partial_\tau \phi_{\Omega_0}}{\partial_\rho \phi_{\Omega_0}}=- \frac{N \omega}{V k}= - \frac{N }{V } c_p,
\end{align}
where we have substituted the definition of phase velocity $c_p=\omega/k$. In the neighborhood of the UH, this becomes
\begin{align}
\label{eq:UH_causal_cone}
    \lim_{N \to 0^+} \frac{{\rm d} \rho}{{\rm d} \tau }\biggl|_\gamma=\lim_{N \to 0^+} - \frac{N \omega}{V k}=1 \,,
\end{align}
where we have used (\ref{eq:hardsol_out}) to evaluate the preferred frequency and momentum of the mode. From \eqref{eq:monochromaric_causal_cone} we can build the effective null coordinate ${\rm d} \Bar{u}= c_p U_\m {\rm d}x^\m + S_\m {\rm d}x^\m $ that, near the UH becomes ${\rm d} \Bar{u}=  {\rm d}\tau - {\rm d} \rho$, so that the ray travels locally at the speed of light in the preferred $(\tau, \rho)$ frame. Taking this into account, the phase ${\cal S}^{\rm red}_0$ becomes:
\begin{align}
\label{eq:pp_action_near_UH_out}
   \mathcal{S}_0^{\rm red} = - \int (\omega N + k V) {\rm d} \tau= - \Omega \tau = - \Omega t + \Omega \int \frac{{\rm d}r }{ N} ,
\end{align}
where we have used the definition of the preferred time coordinate \eqref{eq:preferred_time} in the last step. This shows that constant phase contours in the exterior of the UH correspond to constant $\tau$ surfaces, thus accumulating logarithmically at the UH together with the foliation. 

The counterpart of this mode in the inside region is given by $\psi^{\rm green}$, for which again ${\rm d} \Bar{u}= {\rm d}\tau - {\rm d} \rho$, leading to
\begin{align}
\label{eq:pp_action_near_UH_in}
   \mathcal{S}_0^{\rm green} = - \Omega \Tilde{\tau}  \,,
\end{align}
where $\Tilde{ \tau}$ is the preferred time inside the UH, corresponding to $\Tilde{\tau}(N)=\tau(-N)$, and thus also diverging in the limit $N \to 0^-$.

The presence of logarithmic non-analyticities is a smoking gun of particle emission. This can be understood from quantum tunneling through the horizon -- see \cite{Wilczek00,Srinivasan99, Vanzo11,Giavoni20} for a reference. Although classical trajectories escaping the UH are forbidden, quantum tunneling provides a potential path for particles to escape the gravitational well up to the asymptotic region. The probability for this to happen is computed, in analogy with standard quantum mechanics, through the tunneling rate
\begin{equation}\label{eq:tunneling_rate}
\Gamma= e^{-2{\rm Im}(\mathcal{S}_0)} \,.
\end{equation}
which is determined via the imaginary part of the classical action for the path through the horizon, that acts as an effective potential barrier \cite{Wilczek00}.

For the case at hand, the trajectory that allows modes to tunnel through the UH is given by connecting the green and red trajectories in a single regular path $(t(r),r)$ crossing the horizon. Noting that the Killing time $t$ is real and regular along the UH, and using \eqref{eq:pp_action_near_UH_out} and \eqref{eq:pp_action_near_UH_in}, this leads to
\begin{align}
\label{eq:imS0UH}
  \mbox{Im} (\mathcal{S}_0) = \mbox{Im}\biggl[ \Omega \int_{r_1}^{r_2} \frac{{\rm d}r}{N} \biggr] \,,
\end{align}
with $r_1 < r_{\rm UH} < r_2$. The imaginary part of this integral comes from the simple pole at $N=0$, and can be computed by complexifying the path around $r_{\rm UH}$ by a small imaginary part \cite{DelPorro22_tunneling, Schneider23}. We thus arrive at
\begin{align}
\label{eq:Impart}
  \mbox{Im} (\mathcal{S}_0) = \mbox{Im}\biggl[ \frac{\Omega}{\partial_r N|_{\rm UH}} \int_{r_1}^{r_2} \frac{{\rm d}r}{r-r_{\rm UH}-i \epsilon} \biggr] = \frac{\pi \Omega}{2 \kappa_{\rm UH}} \,,
\end{align}
where we have defined $2 \kappa_{\rm UH} = \partial_r N|_{\rm UH}$ which agrees with \eqref{eq:kappa_UH}. Plugging this into \eqref{eq:tunneling_rate} we obtain the tunneling rate which takes the form of a Boltzmann distribution
\begin{align}
\label{eq:rate_UH}
  \Gamma= \exp \biggl[-\frac{\Omega}{T_{\rm UH}}\biggr],
\end{align}
where the temperature $T_{\rm UH}= \kappa_{\rm UH}/\pi$ is universal and independent on the matter species, and in particular of the exponent $n$ in their dispersion relation.
\subsubsection{Group versus phase velocity}
Let us elaborate on the previous point here. Although the discussion on the specifics of the modes here were done by choosing $n=2$, the tunneling rate $\Gamma$ is actually independent of $n$. This is in contrast to previous results in the literature \cite{Ding16, Herrero-Valea20, DelPorro22}. We trace back this discrepancy to the fact that previous works considered the motion of the mode close to the UH to be controlled by the group velocity of the wave-packet, ignoring its squeezing when $r\rightarrow r_{\rm UH}$. One can immediately recognize that inserting $c_g$ instead of $c_p$ into \eqref{eq:UH_traj} returns ${\rm d}\rho/{\rm d}\tau=n$, giving rise to a different analytical behavior of the action:
\begin{align}
\label{eq:cg_S}
 \mathcal{S}_0= - \int (\omega N + n k V) {\rm d} \tau = - \int ( \Omega + (n-1) kV ) {\rm d} \tau \simeq - (n-1) \int \Lambda \biggl(- \frac{V}{N} \biggr)^{\frac{1}{n-1}} \biggl({\rm d} t - \frac{{\rm d}r}{N} \biggr)\,.
\end{align}
Although this still admits a simple pole for $n=2$\footnote{Note that in \cite{Berglund13} only the case $n=2$ was treated, while in \cite{Ding16} the fractional power was incorrectly handled.}, in general it displays a branch cut with apex on $N=0$, due to the fractional power in $n$. This obstructs the analytic continuation required to make sense of the result. For monochromatic waves with phase velocity $c_p$ instead, the branch cut is absent and only a single pole remains, as we have seen.

The presence of the fractional pole comes directly from \eqref{eq:hardsol_out}, where the hard branches are controlled by the UV anisotropic scaling behavior of the dispersion relation $\omega^2 \sim k^{2n}$. However, this anisotropy is inconsistent with the insertion of $c_g$ in \eqref{eq:cg_S}. In fact, the group velocity implies, within the wave-packet approximation, a linearization of the dispersion relation, akin to the description of a particle following the trajectory given by $\omega= c_g k$. If we trust this approximation down to the UH -- hence ignoring the squeezing of the wave-packet --, we may define the effective null coordinate for a particle following the linearized dispersion relation as
\begin{align}
\label{eq:cg_u}
0= {\rm d} \Bar{u}=n {\rm d} \tau - {\rm d} \rho \simeq (n-1) {\rm d}t - \frac{n}{N} {\rm d}r \,
\end{align}
thus obtaining, for a point particle action defined along this coordinate
\begin{align}
\label{eq:cg_Sg}
\mathcal{S}_g=- \Omega t +   \frac{n}{n-1}  \Omega \int  \frac{ {\rm d}r}{N}  \,.
\end{align}
In this case, one finds the prefactor $n/(n-1)$, thus, reproducing the results in the previous literature and implying a species-dependent temperature. However, as already pointed out, this procedure is incorrect, since it ignores completely the squeezing of the wave-packet as well as the anisotropic behavior that dominates the dynamics of the Lifshitz field at high energies. Once these ingredients are taken into account, one obtains the correct rate \eqref{eq:rate_UH} with a universal temperature. 

As an aside comment, let us note an interesting parallelism with the analog gravity (AG) framework. In analog models, modified dispersion relations are commonplace, and hence the distinction between group and phase velocity becomes substantial. In particular, it is drastically relevant when studying black hole analogs in slow light experiments \cite{Barcelo05}, where it is possible to arbitrarily tune down the group velocity, in order to artificially form a ``group horizon" that may play the role of a causal boundary. Hence, one shall expect particle production. However, this is not the case~\cite{Unruh03,Novello02,Barcelo05}. The production of particles is connected to a mixing of positive and negative frequency modes \cite{ashtekar}, which translates to a condition on the phase velocity instead. Mode mixing is only possible when the phase velocity matches the velocity of the background fluid sustaining the experiment, defining a sort of ``phase horizon". In our case, the analog of this fluid velocity corresponds to the ``aether velocity", with value $-V/N$ \footnote{To see this, it is sufficient to recast the metric into Gullstrand-Painlevé coordinates. See \cite{Parentani15}.}. Indeed, in \eqref{eq:UH_traj} we found that particle production is triggered when $c_p=-V/N$, which happens exclusively at the UH, hence becoming our phase horizon, exactly the same situation as in AG.

Our calculation also saves the thermodynamical picture of the process, since species dependent temperatures allow in principle for the construction of perpetuum mobiles of the second kind \cite{Herrero-Valea20}, unless a UV symmetry is imposed in the dispersion relation. Our results here show that this is unnecessary once the squeezing of the wave-packet is taken into account.

\subsection{Towards the Killing Horizon}

After being produced in the neighborhood of the UH, radiation escapes toward the asymptotic region of large radii. There, observers will measure a wide wave-packet instead of a narrow frequency, which justifies our choice of working with expression \eqref{eq:wavepacket} instead of single wave-modes. The wave-packet will climb the gravitational well following the characteristics of the out-going red mode in Table \ref{t:modes_out}. Its trajectory is determined by its group velocity through the coordinate \eqref{eq:baru} which reads in this case
\begin{equation}\label{eq:dubar}
   {\rm d} \Bar{u}= (c_g(r,\alpha) U_\mu + S_\mu) {\rm d}x^\mu \,.
\end{equation}
The trajectory followed by the wave-packet will actually correspond to level lines of $\bar u$, and hence to ${\rm d}\bar u=0$. From \eqref{eq:classical_action}, this implies ${\rm d} \mathcal{S}_0=k_\mu {\rm d}x^\mu \propto {\rm d} \Bar{u}$. Using $\chi \cdot k= \Omega$ and working in horizon penetrating Eddington-Finkelstein coordinates, we can write
\begin{equation}\label{eq:pp_action_general}
   \mathcal{S}_0=-\int \Omega {\rm d}v - \Omega \int\frac{c_g(r,\alpha) U_r + S_r}{c_g(r, \alpha) U_v + S_v}  {\rm d}r  \,.
\end{equation}
Notice that the coordinate $\bar u$ is clearly $\a$-dependent, i.e. waves with different energy travel along different paths, hence, behaving as accelerated trajectories from a general relativistic perspective. However, let us highlight that \eqref{eq:pp_action_general} does not show any pole structure for $r>r_{\rm UH}$, so that rays can escape up to arbitrary radial values without obstructions by following their classical path.

\begin{figure}
	\includegraphics[scale=0.2]{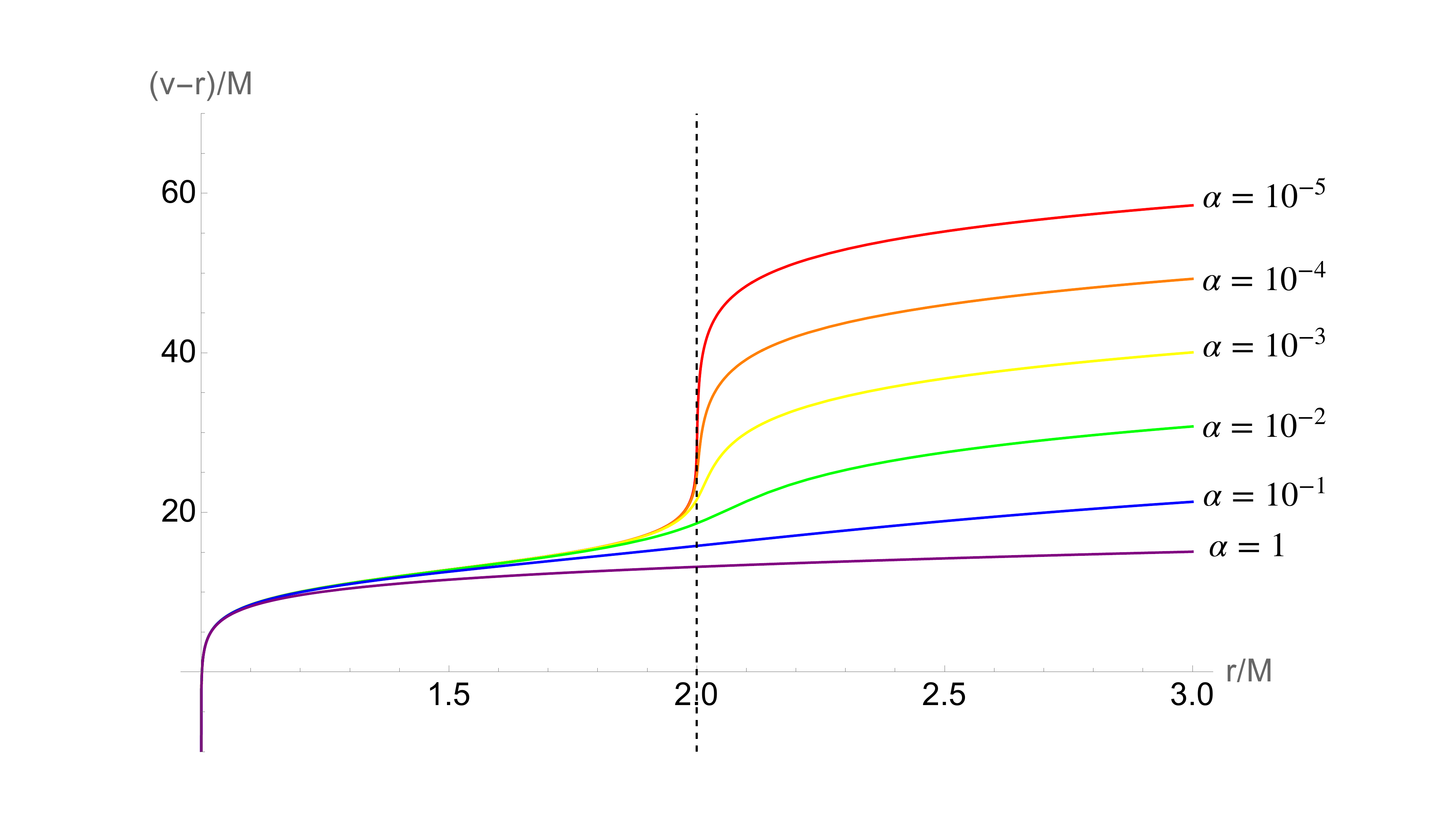}
	\caption{\label{f:lingering} Classical trajectories for $\psi^{\rm red}$ in the $\{t^*,r\}$ plane, evaluated at different values of $\alpha$. Each value of $\alpha$ is reported next to the correspondent ray. We observe a lingering behaviour at the Killing horizon (dashed line) for low-energy modes.}
\end{figure}

After numerically integrating  \eqref{eq:trajectory_vr} in the $\{v,r\}$ plane for different values of $\alpha$ -- see Fig. \ref{f:lingering} --, we observe that the effect of the trajectory's energy-dependence is circumscribed to the surroundings of the Killing horizon, where the rays show a dispersive behavior, typical in modified dispersion relations \cite{Cropp14,Rubio:2023eva}. Indeed, the KH acts as a Newton prism on the wave-packet, separating the different energy modes. This can be easily understood by noting that for $\alpha \rightarrow 0$, the Lorentz violating terms in the action \eqref{eq:lifshitz_action} decouple from the dynamics of the scalar field. This should then correspond to the relativistic limit, where the spectrum of radiation, measured by an observer at infinity, is controlled solely by the Killing horizon, as in GR. In contrast, for $\alpha \rightarrow \infty$ rays cross the whole space-time unaltered, with all their features determined exclusively by the UH. Waves with intermediate values of $\alpha$ must thus interpolate between both behaviors. In particular, we observe that the trajectories linger at the position of the Killing horizon for a time which grows as we decrease $\alpha$.

\subsection{Crossing the Killing Horizon}\label{sec:crossing_KH}

Finally, we complete our description of the radiation's journey up to $\mathscr{I}^+$ by discussing explicitly the modification in the spectrum induced by the presence of the Killing horizon and the dispersion of rays. Let us first point out that although \eqref{eq:dubar} displays no pole at the position of the Killing horizon, we expect to recover the general relativistic result for $\alpha \rightarrow 0$, which corresponds however to particle emission controlled by a pole at $r=2M$ \cite{Wilczek00}, mimicking the result in \eqref{eq:rate_UH}. A remedy to this apparent clash is found by expanding the solution in $k$ for small $\alpha$ around the Killing horizon. From \eqref{eq:krho_eq} we focus on the out-going solution in the region $r>2M$, yielding
\begin{equation}\label{eq:krho_expansion_sol}
   k= \frac{\Lambda r}{r-2M}  \alpha+\frac{ \Lambda  r^3 (M-r)}{2 (r-2M)^4}\alpha^3+{\cal O}(\alpha^4).
\end{equation}

Together with \eqref{eq:cg}, this leads to a group velocity
\begin{equation}\label{eq:cg_expansion}
    c^{\rm red}_g(r,\a)=1+ \frac{3 r^2}{2 (r-2M)^2} \alpha^2 + {\cal O}(\alpha^3) \,.
\end{equation}
which approximates the full result as long as 
\begin{equation}\label{eq:radius_of_convergence}
   \a \ll \frac{r-2M}{r},
\end{equation}
that is, at distances after the lingering has occurred.

At leading order, $c_g$ corresponds to the relativistic value $c_g=1$, but displays a pole at the Killing horizon in the sub-leading correction. Since the full expression is regular at $r=2M$, we conclude that this divergence, despite leading to effective particle emission, must be an artifact of perturbation theory. 
Indeed, we can estimate the validity of the previous expansion by noting that the divergence is actually simply describing the lingering found in Fig. \ref{f:lingering} within the limits of the approximation. A necessary condition for such lingering to exist is a change of concavity in the trajectory, that is
\begin{align}
    \frac{\partial^2 v(r,\alpha)}{\partial r^2}=0.
\end{align}

By differentiating \eqref{eq:trajectory_vr}, this leads to the following condition
\begin{align}
    \frac{\partial^2 v(r,\alpha)}{\partial r^2}=\frac{\partial}{\partial r}\left(\frac{1+c^{\rm red}_g}{N c^{\rm red}_g+V}\right)=0,
\end{align}
whose zeroes can be computed numerically for the full solution. In general, we find two independent solutions, corresponding to the changes of concavity before and after the lingering. However, for a certain critical value $\alpha_{\rm crit}$, these two roots degenerate, indicating an inflection point and the end of the lingering behavior. For the case $n=2$, we numerically estimate the critical value to be
\begin{align}
    \alpha_{\rm crit}\approx 0.114.
\end{align}
The condition $\alpha< \alpha_{\rm crit}$ serves as a rough approximation of the regime of validity of the expansion \eqref{eq:krho_expansion_sol} in small $\alpha$. For values $\alpha> \alpha_{\rm crit}$, the perturbative computation presented here is not valid anymore and one has to consider the full classical action \eqref{eq:pp_action_general} which displays no pole.

The previous expressions describe the behavior of the red mode, while a similar result is attained when examining the orange mode, besides the fact that its support ranges into the interior of the horizon. In particular, its soft branch behaves in the same way as the red mode, with $|c^{\rm orange}_g|$ identical to the absolute value of \eqref{eq:cg_expansion}. This seems to indicate that within this approximation, it is possible to tunnel through the horizon in a quantum mechanically allowed path connecting the orange and red modes, which then behave as Hawking partners in analogy to the standard GR derivation of Hawking radiation \cite{Wilczek00}. Again, this is in close analogy with AG systems, cf. \cite{Coutant12}.

Let us make this last point explicit by considering a path connecting $\psi^{\rm orange}$ and $\psi^{\rm red}$ along their trajectories. Note, however, that the coordinate $t$ in \eqref{eq:pp_action_general} is not regular across the Killing horizon, due to the vanishing norm of $\chi$. Hence, we turn again to horizon penetrating coordinates by changing to $\{ v,r\}$ and choose a regular path $v(r)$ across $r=2M$. Then, the imaginary part of the action along a horizon crossing path is given by
\begin{align}\label{eq:integral_KH}
  \mbox{Im} (\mathcal{S}_0) = -\mbox{Im}\left[\Omega \int_{2M-\delta}^{2M+\delta}\left(   \frac{c_g(r,\alpha) U_r + S_r}{c_g(r, \alpha) U_t + S_t} -\frac{r}{r-2M}\right) {\rm d}r \right],
\end{align}
where $\delta>0$. The result is obtained by the residue of the integrand in \eqref{eq:integral_KH}, which we expand for small $\alpha$ using \eqref{eq:krho_expansion_sol}, so that
\begin{equation}\label{eq:Impart_KH}
      \mbox{Im} (\mathcal{S}_0) = \mbox{Im}\left\{ \int_{2M- \delta}^{2M+ \delta} \Omega \biggl[\frac{4
   M-12 \alpha ^2 M}{r-2M }-\frac{24 \alpha ^2 M^4}{(r-2M)^4} - \frac{48 \alpha ^2
   M^3}{(r-2M)^3} - \frac{36 \alpha ^2 M^2}{(r-2M)^2} + {\cal O}(\alpha^3) \biggr]  \, {\rm d}r \right\}\,.
\end{equation}

Note that this integral only admits a pole at the precise position of the Killing horizon. To extract explicitly its imaginary part, we simply complexify the integration path around $r=2M$ by replacing $2M \to 2M+ i \epsilon$, with $\epsilon>0$, and taking the limit $\epsilon \to 0$ afterwards. Doing this, one can easily prove that the higher order poles lead to
\begin{equation}
      \int_{2M- \delta}^{2M+ \delta} \frac{1}{(r-2M)^m} \, {\rm d}r \propto \frac{1}{\delta^{m-1}} \qquad \mbox{for} \quad m>1
\end{equation}
which is explicitly real. The only imaginary contribution is given by the simple pole in \eqref{eq:Impart_KH}, arriving at
\begin{equation}\label{eq:Impart_KH_2}
      \mbox{Im} (\mathcal{S}_0)=\mbox{Im} \left\{\int_{2M- \delta}^{2M+ \delta} \Omega \biggl[\frac{4
   M-12 \alpha ^2 M}{r-2M - i \epsilon} \biggr]  \, {\rm d}r\right\} = \frac{\Omega \pi}{\kappa_{\rm KH}}(1-3\alpha^2) \,,
\end{equation}
where we have introduced the surface gravity of the Killing horizon $\kappa_{\rm KH}= (4M)^{-1}$. The tunneling rate is then
\begin{equation}\label{eq:tunneling_rate_KH}
    \Gamma=\exp \biggl[-\frac{2\pi\Omega}{\kappa_{\rm KH}} (1-3\alpha^2)\biggr] \,,
\end{equation}
from which we can extract an $\alpha$-dependent notion of `temperature' 
\begin{equation}\label{eq:temperature_KH}
     T_\a = \frac{\kappa_{\rm KH}}{2 \pi (1-3\alpha^2)}= \frac{T_{\rm H}}{1-3\alpha^2}=T_{\rm H}\left(1+3\alpha^2+{\cal O}(\alpha^3)\right) \,.
\end{equation}
with $T_{\rm H}=\kappa_{\rm KH}/2\pi$ being the Hawking temperature of the black hole as computed in GR, or equivalently in the decoupling limit. 

Let us comment a bit on this result. Earlier, we discussed how the trajectories of the red and orange mode become reminiscent to the Hawking partners in GR within the limit $\alpha\rightarrow 0$. The only difference between them comes from the fact that the red mode carries a positive Killing energy $\Omega$ and escapes the gravitational well, while the orange mode contains negative energy and falls into the singularity, peeling from the interior of the horizon instead. Such image is consistent with our findings in \eqref{eq:deltaroot}, which show that the turning point of the orange mode is always behind the Killing horizon. This leads to the existence of a tunneling path that connects both modes and can be interpreted as a single ray consisting of the complex conjugate of the orange mode coming from the inside of the horizon towards its exterior -- thus carrying positive Killing energy --, and tunneling out via a quantum mechanical process.

This yields the tunneling rate \eqref{eq:tunneling_rate_KH} which is controlled by the quantity $T_\a$. Although we call it temperature, the energy dependence within implies that this is not, strictly speaking, a temperature. The Lorentz violating operators in the dispersion relation induce a small deviation from the thermal behavior assigned to the Killing horizon in the GR setting. An observer at infinity would measure a deviation from the expected thermal spectrum at low energies $\a < \a_{\rm crit}$, while recovering a perfect black body spectrum at high energies, controlled by $T_{\rm UH}$. 

As a side comment let us look at this result from a different viewpoint. Albeit being an incorrect procedure close to the UH, we can nevertheless introduce the group velocity $c_g$ as a way to linearize the dispersion relation \eqref{eq:phi_eom} as $\omega = c_g k$, condensing all information about the superluminal propagation into $c_g$. This is equivalent to defining a refractive index in a dispersive medium -- the dispersion relation is non linear, but the motion at low-frequencies is effectively described by a linearized version thereof, with a pointwise dependent coefficient, the refractive index \cite{Barcelo05}. Together with this concept, one can also show that there exists a critical frequency, the so-called plasma frequency, at which this approximation breaks down and the UV behavior of the high-frequency modes must be described by the full dispersion relation. Along this line, we can think about $\a_{\rm crit}$ as a plasma frequency, below which the evolution of modes is governed by the effective description given by $c_g$, feeling the presence of the KH. UV modes, instead, do not enjoy the same features, cross freely the Killing horizon, and obey the high-energy part of the dispersion relation instead, which is sensible to the UH only. An alternative derivation of $T_\alpha$, based on this approach, is provided in appendix \ref{app:KH_T}.

Let us also point out that within the Wilsonian picture of effective field theories, and since we compute it in perturbation theory in $\alpha$, the correction obtained here should be universal, as long as the dispersion relation contains a $k^4$ term, even if higher powers of the momentum are present. The only difference that we might expect in the general case is a small ${\cal O}(1)$ correction in the coefficient accompanying $\alpha^2$, due to a different value of $\beta_4$, which has been fixed to unity here. Nonetheless, the general result can be easily obtained by rescaling $\Lambda$, which amounts to replacing $\alpha^2 \rightarrow \beta_4 \alpha^2$ in the final result.


\section{Quantum state}\label{s:QS}

In the previous sections we discussed how the behavior of the modes shows a mode mixing -- hence particle production -- while we perceive a sort of a reprocessing of the out-going radiation at the KH. Note that in doing so, we assumed implicitly the regularity of the underlying quantum state at both horizons. If this were true, it would in turn imply the existence of a globally defined vacuum state which is Unruh-like at the UH, and mimics the standard Unruh vacuum state for freely falling (geodesic) observers at the KH, at least for low energy rays. Any deviation will either result in a divergence at the UH, or in a firewall -- with typical energies of order $\Lambda$ -- at the KH \cite{Goto18}. In order to justify our approach, let devote some space to study this issue in greater detail.

\subsection{LV-Unruh state}

By working in the tunneling picture, we arrived to eq.\eqref{eq:rate_UH}, describing the tunneling amplitude for a particle crossing the UH. However, the construction of the path through $r=r_{\rm UH}$ allowing for this can as well be understood as providing an analytical continuation of $\psi^{\rm red}$ through the UH. Indeed, $\psi^{\rm red}=\psi^{\rm red}_0 \, e^{i \mathcal{S}_0}$, with $\mathcal{S}_0$ given by \eqref{eq:pp_action_general}, only has support for $r>r_{\rm UH}$ and cannot be analytical in the boundary of this region \cite{Jacobson03}. Close to the UH, the radial dependence of the mode leads to a branch-cut in $\mathcal{S}_0$, produced by the logarithmic accumulation of the phase contours
\begin{align}\label{eq:log_dep_red}
   \psi^{\rm red} \propto \theta(r-r_{\rm UH}) \exp \biggl[ i\frac{\Omega}{2 \kappa_{\rm UH}} \int^r \frac{{\rm d}r}{r-r_{\rm UH}} \biggr]=\theta(r-r_{\rm UH}) \exp \biggl[ i\frac{\Omega}{2 \kappa_{\rm UH}} \log (r-r_{\rm UH}) \biggr] \,,
\end{align}
where $\theta(r-r_{\rm UH})$ is the Heaviside distribution. Analogously, the green mode's support is restricted to the region $r<r_{\rm UH}$
\begin{align}\label{eq:log_dep_green}
   \psi^{\rm green} \propto \theta(r_{\rm UH}-r) \exp \biggl[- i\frac{\Omega}{2 \kappa_{\rm UH}} \int^r \frac{{\rm d}r}{r-r_{\rm UH}} \biggr]=\theta(r_{\rm UH}-r) \exp \biggl[ - i\frac{\Omega}{2 \kappa_{\rm UH}} \log (r_{\rm UH}-r) \biggr] \,,
\end{align}
where we have taken $\Omega>0$ for both modes. 

The analytic continuation of $\psi^{\rm red}$ into the region $r<r_{\rm UH}$ is obtained in the standard way, through the corresponding continuation of the complex logarithm \cite{Jacobson03}. Depending on whether we place the branch-cut in the upper or lower-half of the complex plane, we have the following two possible continuations, distinguished by a $\pm$
\begin{equation}
\label{eq:analytical_continuation}
\begin{split}
   \left.\psi^{\rm red}\right|_{r<r_{\rm UH}} \propto \theta(r_{\rm UH}-r) \exp \biggl[ i\frac{\Omega}{2 \kappa_{\rm UH}} \log (|r-r_{\rm UH}|)  \pm \frac{\Omega \pi}{2 \kappa_{\rm UH}}\biggr]  \propto e^{ \pm \frac{\Omega \pi}{2 \kappa_{\rm UH}}} (\psi^{\rm green})^*\,.
\end{split}
\end{equation}

Therefore, by gluing $\psi^{\rm red}$ with its two possible analytic continuations, we can construct two functions that are analytical through $r=r_{\rm UH}$ in one of the two halves of the complex $r-$plane
\begin{align}\label{eq:pos_neg_dec}
   \Psi^\pm_\Omega=\mathcal{C}^\pm \biggl[\psi^{\rm red} + e^{ \mp \frac{\Omega \pi}{2 \kappa_{\rm UH}}} (\psi^{\rm green})^* \biggr]\,,
\end{align}
where the subscript $\Omega$ has been reintroduced to stress the fact that the mode functions are computed at fixed $\Omega>0$. 

By construction, analyticity in the lower (upper) half of the complex plane implies that the function contains only positive (negative) energy modes -- see e.g. \cite{Jacobson03} --, and hence the modes $\Psi_{\Omega}^{\pm}$ must be orthogonal when the sign differs. This leads to the normalization conditions $\langle \Psi^+_\Omega,\Psi^+_{\Bar{\Omega}} \rangle = \langle \Psi^-_\Omega,\Psi^-_{\Bar{\Omega} }\rangle= \delta_{\Omega,\Bar{\Omega}}$ and $\langle \Psi^+_\Omega,\Psi^-_{\Bar{\Omega}} \rangle =0$ -- see appendix \ref{a:inner_product} for further details -- which imply
\begin{align}\label{eq:norm_coeff}
   \mathcal{C^\pm}= \frac{1}{1-e^{ \mp \frac{\Omega \pi}{ \kappa_{\rm UH}}}}\,.
\end{align}

As another consequence of analyticity, the modes $\Psi_\Omega^\pm$ can be decomposed in terms of purely positive or negative Minkowski modes $f_\Omega^{\pm}$, defined as solutions of the field equations in the asymptotic region\footnote{This means that, by definition, $f_{\Omega}^\pm$ is a solution of \eqref{eq:asympt_eom}, and satisfies $\chi^\m \partial_\m f_\Omega^\pm=\mp i \Omega f_\Omega^\pm$.}, where the space-time is asymptotically flat
\begin{align}\label{eq:bogolyubov}
   \Psi^+_\Omega= \int_0^{+ \infty} {\rm d} \Bar{\Omega} \, \alpha_{\Omega \Bar{\Omega}} f^+_{\Bar{\Omega}} \,, \qquad \Psi^-_\Omega= \int_0^{+ \infty} {\rm d} \Bar{\Omega} \, \beta_{\Omega \Bar{\Omega}} f^-_{\Bar{\Omega}} \,.
\end{align}
The set $\{f_\Omega^{\pm}\}$ forms a complete basis for analytic functions with support on the whole real line for $r$, and as such the Bogolubov coefficients $\alpha_{\Omega \Bar{\Omega}}$ and $\beta_{\Omega \Bar{\Omega}}$ then describe the particle content of $\psi^{\rm red}$ in terms of positive and negative frequency modes, respectively. Indeed, one can check explicitly that $\psi^{\rm red}= \Psi^+_\Omega + \Psi^-_\Omega$, from which
\begin{align}\label{eq:bogolyubov_psired}
   \psi^{\rm red}= \int_0^{+ \infty} {\rm d} \Bar{\Omega} \, (\alpha_{\Omega \Bar{\Omega}} f^+_{\Bar{\Omega}} + \beta_{\Omega \Bar{\Omega}} f^-_{\Bar{\Omega}}) \,.
\end{align}

Provided with a complete basis of operators, we are well-equipped to face the issue of the quantum state carried by the radiation within our setting. In the asymptotically flat region, we define the standard Minkowski vacuum $| 0_{\rm M} \rangle$, through
\begin{align}
   a_{\Omega} |0_{\rm M}\rangle=0,
\end{align}
where $a_{\Omega}$ is the operator associated to the annihilation of a positive mode $f^+_\Omega$. An analogous definition at the UH, employs then the adiabatic basis $\{ \psi^{\rm red}_\Omega \}$ which spans what we call the ``LV-Unruh" state $|0_{\rm UH}\rangle$
\begin{align}\label{eq:b_Omega}
   b_\Omega |0_{\rm UH}\rangle =0,
\end{align}
with $b_\Omega$ the annihilation operator for $\psi^{\rm red}_\Omega$. This state describes a vacuum for a freely falling observer that crosses the UH and measures no $\psi^{\rm red}$ particles.

Let us now consider the global state of the system $|\bar 0 \rangle$ and enforce it to agree with $|0_{\rm M} \rangle$ on $\mathscr{I}^-$ -- due to suitable initial conditions -- and with $| 0_{\rm UH} \rangle$ at the UH, for the sake of regularity. After the radiation, emitted from the UH, has escaped to $\mathscr{I}^+$, we can examine the fate of $|\bar 0 \rangle$ by evaluating the expectation value of the number operator $\hat{N}_\Omega=b^\dagger_\Omega b_\Omega$ with respect to $|0_{\rm M}\rangle$ \cite{birrell_davies_1982}. Therefore
\begin{align}\label{eq:particle_number}
  \langle \hat{N}_\Omega \rangle_{\rm M}=\langle0_{\rm M}| b^\dagger_\Omega b_\Omega|0_{\rm M}\rangle = \int {\rm d}\Bar{\Omega} \, |\beta_{\Omega \Bar{\Omega}}|^2 \,.
\end{align}
By definition $\beta_{\Omega \Bar{\Omega}}=- \langle \psi^{\rm red},f^-_{\Bar{\Omega}} \rangle$, and thus
\begin{align}
  \mathcal{C}^-=\langle \Psi^-_\Omega,\psi^{\rm red}_{\Omega} \rangle = \int {\rm d} \Bar{\Omega} \beta_{\Omega \Bar{\Omega}} \langle f^-_{\Bar{\Omega}},\psi^{\rm red}_{\Omega} \rangle=-\int{\rm d} \Bar{\Omega} \beta_{\Omega \Bar{\Omega}} \beta^*_{\Omega \Bar{\Omega}}=-\int {\rm d} \Bar{\Omega} |\beta_{\Omega \Bar{\Omega}}|^2\,,
\end{align}
which leads eventually to
\begin{align}\label{eq:spectrum_UH}
   \langle \hat{N}_\Omega \rangle_{\rm M}=\int {\rm d}\Bar{\Omega} \, |\beta_{\Omega \Bar{\Omega}}|^2=-\mathcal{C}^-= \frac{1}{e^{  \frac{\Omega }{ T_{\rm UH}}}-1} \,.
\end{align}
Hence, we discovered a Bose-Einstein distribution of particles with temperature $T_{\rm UH}=\kappa_{\rm UH}/\pi$, implying that $|\bar 0 \rangle$ corresponds to a non-empty state filled with a thermal distribution of $\psi^{\rm red}_{\Omega}$ particles at $\mathscr{I}^+$, due to particle production by the horizon. This reciprocates the standard situation in GR, where the state is totally fixed by the non-analyticity present in the modes -- in this case $\psi^{\rm red}$ --, and the constraint to be regular at the horizon.

\subsection{Quantum state at the KH}
Although the discussion in the previous section seems somewhat rigorous, it has ignored the presence of the KH. Indeed, in the decoupling limit the dynamics of the out-going radiation is fully determined by the KH, and the quantum state must then match the standard Unruh vacuum at horizon crossing. For non-vanishing values of $\alpha$ we wonder what is the role of the KH, and if its presence modifies our previous conclusions in a significant manner.

Close to $r=r_{\rm KH}$, the $\psi^{\rm red}$ modes follow the effective null coordinate $\bar u$, defined in \eqref{eq:dubar}, which depends explicitly on the value of $\alpha$. For $\a \ll 1$, their evolution mimics that of a relativistic ray in GR, peeling from the KH for sufficiently long time. This produces a large blueshift that is responsible for its adiabatic behavior. Hence, the mode can be described in the WKB approximation which allows for a simpler definition of quantum states in terms of oscillators. This is indeed the assumption underlying the description in the previous section, and the definition of $|0_{\rm UH} \rangle$.

Then, by assuming that the WKB condition is valid in the neighborhood around the KH, at least for $\alpha < \alpha_{\rm crit}$, we follow the same steps that we performed in the previous section, simply by replacing the UH by the KH. Note, however, that we tacitly assume $|\bar 0\rangle$ corresponds to the vacuum state -- hence to a standard Unruh state -- at KH crossing. We will discuss whether or not this remains compatible when contrasted with the assumptions performed close to the UH; for a moment we will assume compatibility. As discussed in subsection \ref{sec:crossing_KH}, we connect the orange and red mode with a regular path through the KH (for small $\alpha$ modes) which then leads to a basis of analytic modes\footnote{Note that the tail of higher order poles in \eqref{eq:Impart_KH} is irrelevant here, since only the imaginary part of the exponent is required to fix the analytic continuation}
\begin{align}
     \Phi^\pm_\alpha=\mathcal{A}^\pm_\a \biggl[\psi^{\rm red}_\a + e^{ \mp \frac{\Omega }{2 T_\a} }(\psi^{\rm orange}_{\rm soft})_\a^* \biggr]\,,
\end{align}
where $T_\a$ is given in \eqref{eq:temperature_KH}. The corresponding spectrum measured by an observer on $\mathscr{I}^+$ is then
\begin{align}\label{eq:N_a}
     \langle \hat{N}_\a \rangle_{\rm M}=\langle0_{\rm M}| c^\dagger_\a c_\a|0_{\rm M}\rangle = \int {\rm d}\Bar{\Omega} \, |\beta_{\Omega \Bar{\Omega}}^\a|^2=-\mathcal{A}^-_\a= \frac{1}{e^{  \frac{\Omega }{ T_{\a}}}-1} \,,
\end{align}
where $c_\a$ is the annihilation operator associated to $\psi^{\rm red}_\a$ for small $\alpha$ in the vicinity of the KH. In this case, this is a ``quasi-Bose-Einstein" distribution, displaying deviations from thermality of order $\alpha^2$.

Note that, as expected, \eqref{eq:N_a} reduces to the relativistic result in the limit $\a \to 0$, becoming exactly thermal with temperature $\kappa_{\rm KH}/2 \pi$. This is a strong hint that a freely falling (geodesic) observer will see almost vacuum whenever the Lorentz violation remains small.

\subsection{States compatibility}
What is left to complete our argument, is to confirm the compatibility of the assumptions made along the  previous sections for the state $|\bar 0\rangle$. In fact, we need to show that $|\bar 0\rangle$ corresponds to a vacuum at both horizons, UH and KH, -- for the latter at least for small $\alpha$.

Although there is no real non-analyticity at the KH for arbitrary values of $\alpha$, and thus there is not real need to enforce $|\bar 0\rangle$ to be vacuum there; any other choice would lead to strong deviations from thermality. In particular, the presence of low energy rays with $\alpha<\alpha_{\rm crit}$ lingering in the area around the KH could lead to the creation of a firewall at energy scale $\sim \alpha_{\rm crit}\Lambda $ for any observer who falls into the gravitational well and does not measure vacuum at the KH. Hence, healthy thermodynamics and regularity of space-time seem both to require vacuum conditions at the horizons.

It should be mentioned that the definition of vacuum relies strongly on the existence of an adiabatic basis of the solution space ($\{ \psi^{\rm red}_\Omega \}$ in our case) that allows to define a set of ladder operators in the corresponding Fock space. At the UH, this is provided by the operators $\{b_{\Omega}\}$ in \eqref{eq:b_Omega} for all $\Omega$, thanks to the infinite blue-shift of the modes. At the Killing horizon, however, the same holds true only in the low energy part of the spectrum, for which it is the lingering behavior of the rays what allows to satisfy the adiabaticity condition in the exterior of the KH at $\alpha\ll 1$.

The last remaining question is whether the rays $\psi^{\rm red}_\Omega$ satisfy the adiabaticity condition also in the intermediate region between the two horizons. If the answer is in the positive, then the adiabatic basis $\{\psi^{\rm red}_\Omega \}$ and its correspondent set of creation and annihilation operators is sufficiently well-defined in the whole region. This means that we can evolve the ladder operators, mode by mode, without mixing between positive and negative norm modes along the trajectories \cite{agullo2015preferred}. Equivalently, we can say that in the low-$\a$ part of the momentum space, the state at the KH and the one at the UH are connected by a unitary operator $\mathcal{U}_\a$
\begin{align}
    |\bar 0 \rangle_{\rm KH} = \mathcal{U}_\alpha |0\rangle_{\rm UH} \,.
\end{align}
Intuitively speaking, validity of the WKB condition in the intermediate region implies that one produces a quantum described by $\psi^{\rm red}_\Omega$ at the UH and evolve it along its trajectory adiabatically up to the KH without additional particle production. 

Let us close the argument and analyze whether or not the WKB condition is fulfilled. As a first step, we rewrite the group velocity \eqref{eq:cg} as
\begin{align}
     c_g(r,\a)= \frac{\partial \omega}{\partial k}= \frac{\partial_r \omega}{\partial_r k}= \frac{ \omega'}{ k'}\,,
\end{align}
where a prime denotes the $r$-derivative. Differentiating \eqref{eq:KE} with respect to $r$ yields
\begin{align}
     0=(\omega + k) N' + N \omega' + V k'\,,
\end{align}
where we used that $N'=V'$, which holds true in our case. Combining the previous two expressions we thus get
\begin{align}
    \omega' = - \frac{c_g N' ( \Omega + k )}{N(c_g N + V)} \implies \frac{\omega'}{\omega^2}=- \frac{c_g N N' (\Omega + k)}{(\Omega-kV)^2(c_gN+V)}\,.
\end{align}

In order to properly evaluate the WKB condition we must assign a meaning to the ``dot derivative" in $| \Dot{\omega}| \ll \omega^2$. This is, as underlined in \cite{Schutzhold13}, an observer-dependent statement. More precisely, it is strongly affected by the choice of the clock that measures the variation of $\omega$. Following the previous discussions, we choose a low energetic in-falling observer, whose trajectory coincides with a low energy trajectory of $\psi^{\rm blue}$. For small $\alpha$, the world-line of our observer will be very close to those of a GR observer \eqref{eq:bluemodetraj}. We choose to label the trajectory using the $\Bar{u}$ coordinate for the out-going red mode, that thus defines our clock -- see appendix \ref{a:ETF} for details. This yields
\begin{align}\label{eq:omegadot}
    \dot{\omega} = \frac{{\rm d} \omega(r(\Bar{u}),\a)}{{\rm d} \Bar{u}} = \frac{{\rm d} r(\Bar{u})}{{\rm d} \Bar{u}} \omega'\,,
\end{align}
where $\partial_{\Bar{u}}r$ is given in \eqref{eq:labelubar}. Substituting it into \eqref{eq:omegadot} and dividing by $\omega^2$, we get
\begin{align}\label{eq:adiabatic_ratio}
   \frac{ \dot{\omega} }{ \omega^2}  = \frac{{\rm d} r(\Bar{u})}{{\rm d} \Bar{u}} \frac{ \omega'}{\omega^2} =- \biggl( \frac{c_g U_r + S_r}{c_g N + V}  - \frac{r}{r-2M} \biggr)^{-1} \frac{c_g N N' (\Omega + k)}{(\Omega-kV)^2(c_gN+V)}\,.
\end{align}

\begin{figure}
	\includegraphics[scale=0.4]{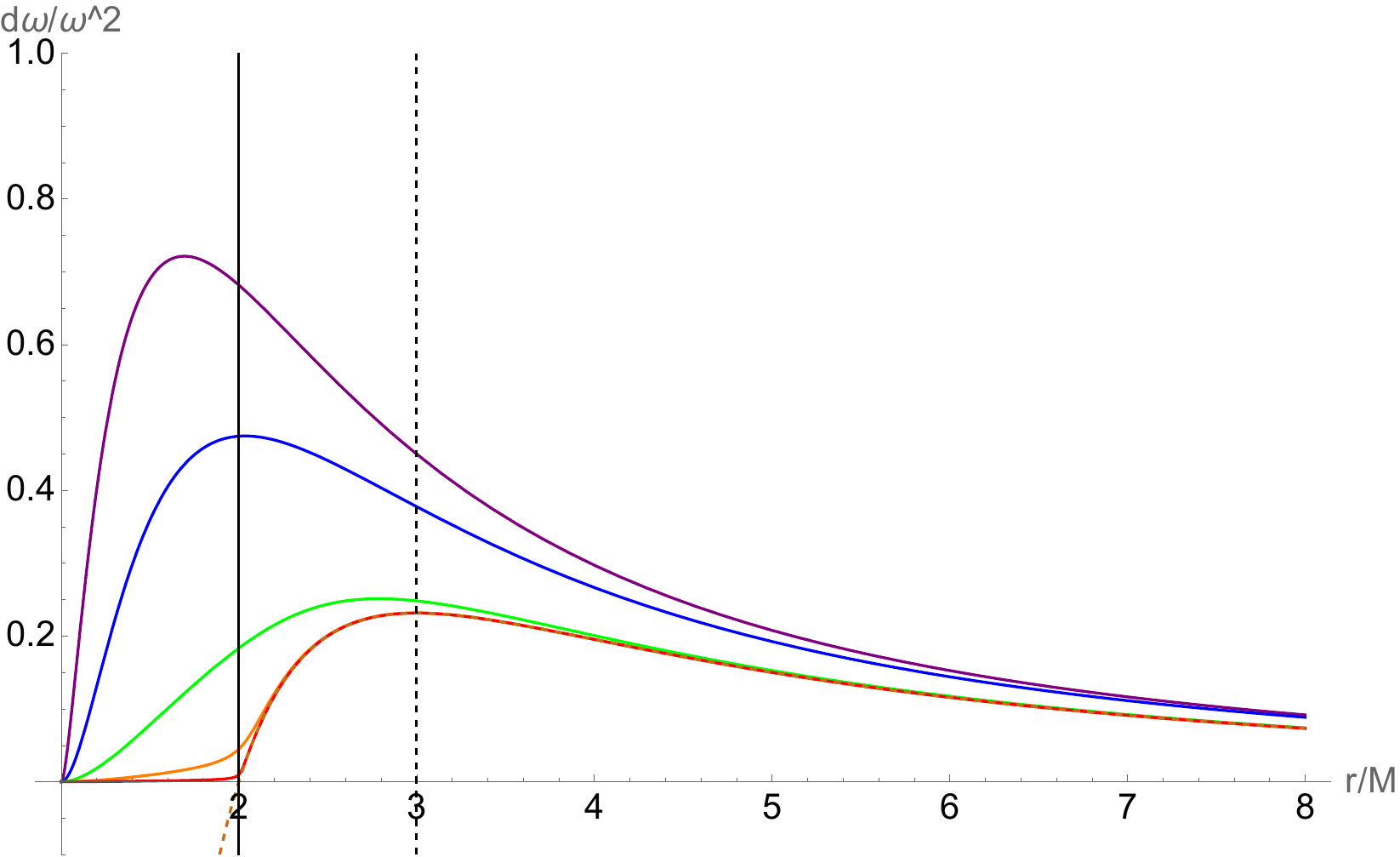}
	\caption{\label{f:WKBcond} Adiabatic condition for modes $\psi^{\rm red}_\Omega$ with different $\alpha$: $\a=10$ (purple), $\a=1$ (blue), $\a=10^{-1}$ (green), $\a=10^{-2}$ (orange), $\a= 10^{-3}$ (red). The dashed orange line represent the checking of the WKB condition for the relativistic case. The vertical lines are $r=2M$ (solid) and $r=3M$ (dashed). All lines are computed with $\Omega= 1.6 \times 10^{-1}$, laying close to the peak of the emission from the UH, which sits at $\Omega= 1/ (2\pi M)$.}
\end{figure}

As it can be seen in Fig. \ref{f:WKBcond}, in between the two horizons the ratio $|\dot \omega / \omega^2|$ decreases as we lower $\alpha$, which corresponds to the region where problems could arise due to apparent effective particle production at the KH. This is simply a consequence of the fact that the hard solutions in \eqref{eq:krho_eq} behave as $k^{\rm red} \sim 1/ \a$ for $n=2$ and in the region between the two horizons -- see appendix \ref{a:pert_hard} for further details. Consequently, one also has $c_g^{\rm red} \sim 1/ \a $ in between both horizons, which leads to
\begin{align}\label{eq:wkb}
   \lim_{\a \to 0^+} \frac{ \dot{\omega} }{ \omega^2} =0 \quad \mbox{for} \quad r \in [M, 2M]\,.
\end{align}
Therefore, we can conclude that the WKB approximation and hence adiabaticity, hold true in the intermediate region, being more exact as we lower $\alpha$. This leads us to conclude that the existence of a single state $|\bar 0\rangle$, which is vacuum at both horizons, is possible. Moreover, let us note from Fig. \ref{f:WKBcond} that for $\alpha \rightarrow 0$, the shape of $ |\dot{\omega} / \omega^2| $ recovers the GR result, with the usual peak around $r=3 M$, thus underlying the breakdown of the WKB approximation in the light ring. Our results here are in agreement with previous studies \cite{Schutzhold13}, where it has been shown that the WKB approximation holds all the way through until the KH.

\section{Discussion}

In this work, we have studied the thermal properties of black holes in Lorentz violating gravity. More precisely, we have focused on the emission and subsequent propagation of Hawking radiation on metric-aether, spherically symmetric, black holes, associated to a field with a modified (superluminal) dispersion relation, in a setting where the background gravitational dynamics is provided by the low energy limit of Ho\v rava gravity -- also known as khronometric theory. 
The key novelty of this framework with respect to similar investigations in analogue gravity \cite{Barcelo05} is provided by the presence of an universal horizon -- a trapping surface for signal of arbitrary speed located behind the Killing horizon.

We have seen that the Universal horizon, being here a true causal horizon, plays a fundamental and dual role -- the requirement of regularity at its surface fixes unambiguously a globally defined Unruh-like vacuum state associated to observers falling with the preferred (aether) frame. Once this state is fixed, it is possible to prove that high energy modes ($\alpha>\alpha_{\rm crit}$) arrive to the asymptotic region in a thermal ensemble with a universal temperature given by $T=\kappa_{\rm UH}/\pi$. Two key ingredients contribute to this result: 1) the reversal of the lapse function's sign inside the UH (overlooked in early works on the subject) is crucial in the interpretation of the Hawking partner (green mode in our nomenclature) as a negative energy mode moving from the UH to the central singularity; 2) the squeezing and blueshift of the wave-packets that turns them into monochromatic waves at the UH, so that only the signal velocity (infinite momentum limit of the phase velocity) does matter for their propagation. This is also true at a Killing horizon in the relativistic case, but less appreciable due to the degeneracy between phase and group velocities for linear, relativistic, dispersion relations in the high energy limit. We have also seen that this second point is crucial in eliminating any species dependence -- the $(n-1)/n$ prefactor found in several previous works \cite{Herrero-Valea20,DelPorro22,DelPorro22_tunneling}, hence avoiding any possibility for violations of the second law of thermodynamics -- see discussion in \cite{Herrero-Valea20}. 

This would be per se a very compelling picture, but unfortunately (or luckily?) the journey of the superluminal modes leaving the exterior of the UH is far from trivial. Indeed, we have also found that the KH acts as a Newton prism, separating modes of different Killing frequency. Rays linger at the KH for a time depending of their energy --more lingering for lower $\alpha$'s. As said also above, such prism has the analogue of a plasma frequency, which can be identified with $\Omega\sim \alpha_{\rm crit}\Lambda$, beyond which the KH plays no effect on the passing rays.

Of course, such an effective refractive index has a very relevant effect on the outgoing radiation which in all its complexity it is difficult to grasp, especially if one evolves wave packets containing modes of different frequencies. We have nonetheless used the fact that, at lowest order, outgoing rays for which $\alpha\ll\alpha_{\rm crit}$ show an effective pole in their propagation, and have a companion behind the KH very closely mimicking the usual trajectory of the Hawking partner in the relativistic limit ($\Lambda\to\infty$). This allows to perform a tunneling calculation similar to the general relativistic case, but taking into account the first leading correction due to a non-vanishing $\alpha$. Our result shows that radiation with $\alpha\ll\alpha_{\rm crit}$ is reprocessed at the KH in such a way that their contribution at infinity corresponds to a (quasi) thermal spectrum with a weakly energy dependent temperature $T_\a=T_{\rm H}\left(1+3\alpha^2+{\cal O}(\alpha^3)\right)$ -- see Eq.~\eqref{eq:temperature_KH}. This is quite striking and in full agreement with the common wisdom acquired e.g.~in the analogue gravity community, where very similar calculations at KHs -- but in the absence of UHs -- have been performed (see e.g.~\cite{Barcelo05} and references therein).

The presence of a component of radiation dominated at low energies by the KH opens a problem of compatibility. We have fixed the state to be an Unruh state for preferred frame observers -- vacuum at past null infinity and at UH crossing in the aether frame --, but the low energy result implicitly assumes the usual Unruh state defined with respect to metric freely falling observers at the KH. An incompatibility between these two states could imply the presence of a firewall at energy $\mathcal{O}(\Lambda)$ at the KH. We have investigated this issue and showed that for low energy modes - those affected by the KH -- the WKB approximation is perfectly valid from the UH till well beyond the KH. Hence the vacuum defined by the regularity requirement at the UH is still valid at the KH.

Overall, the picture emerging from all of these pieces is remarkable consistent. A black hole in Lorentz violating gravity, if endowed with a UH, and of mass $M>\Lambda$, will radiate with a spectrum which for Killing frequencies below $\alpha_{\rm crit}\Lambda$ -- in this case those accounting for the peak of the spectrum and the low energy tail -- is dominated by the KH surface gravity, but which also displays a high energy Planckian tail characterized also by $\kappa_{\rm UH}$. As the black hole evaporates, its Hawking temperature will increase until it reaches $T_{\rm UH}\simeq \alpha_{\rm crit}\Lambda$ -- corresponding to a mass of order $M \simeq M_p^2/(\Lambda \a_{\rm crit})$-- when its spectrum will start to be dominated by the UH surface gravity. Accordingly, while in the initial stages of the evaporation the emitted radiation will appear to emanate from the KH, at late stages it will appear to originate from the UH. So, while initially it will be only possible to probe with Hawking radiation the region outside the KH, at later times the whole space-time outside the UH will be accessible to distant observers. The true causal structure of these black holes will unveil only at the end of their life-time.

In conclusion, we think that this work should be taken as further evidence for the resilience of horizon thermodynamics in gravitational theories. Indeed, the lessons gathered here should be of interest for  a broader community than that working on  Lorentz violating gravity. The survival of black holes, and of Hawking radiation, in settings entailing signals of arbitrary high speed is far from trivial, and hints again to a very deep connection of gravity and thermodynamics beyond GR. Also, we would like to stress how these more general settings for studying well known phenomena like Hawking radiation can be extremely useful in teaching lessons sometimes hidden by the high degree of symmetry of the relativistic framework. 

Of course we do not deem this investigation definitive. A full spectrum is still to be derived, probably with the help of semi-analytical or full fledged numerical methods. Also, this is just a first step in understanding the thermodynamics of black holes in Lorentz violating theories, in particular if the four laws of black hole thermodynamics can be extended to them (see e.g.~\cite{Pacilio:2017emh,Pacilio:2017swi}). In this sense, it would be nice first to have rotating solutions as well, however to date no (non-slowly) rotating black holes with a UH are known, except for a BTZ black hole in $(1+2)$-dimensions~\cite{PhysRevD.90.044046}.  We hope to tackle these and several other open puzzles in the future, and that this investigation will stimulate others to do the same.


\acknowledgements
We are grateful to Ted Jacobson, David Mattingly, Thomas Sotiriou, Sergey Sibiryakov and Enrico Barausse for insightful discussions during the preparation of this work. The work of F. D. P., S. L., and M. S. has been supported by the Italian Ministry of Education and Scientific Research (MIUR) under the grant PRIN MIUR 2017-MB8AEZ. The work of M. H-V. has been supported by the Spanish State Research Agency MCIN/AEI/10.13039/501100011033 and the EU NextGenerationEU/PRTR funds, under grant IJC2020-045126-I; and by the Departament de Recerca i Universitats de la Generalitat de Catalunya, Grant No 2021 SGR 00649. IFAE is partially funded by the CERCA program of the Generalitat de Catalunya.

\appendix

\section{Killing Horizon temperature} \label{app:KH_T}

In the main text we obtained the effective temperature of the KH \eqref{eq:temperature_KH} by means of the tunneling method. However, in this appendix we provide two alternative derivations of this result, based on constructing effective group velocity horizons, and on the definition of an effective temperature function.

\subsection{Disformal transformation}

\begin{figure}
	\includegraphics[scale=0.3]{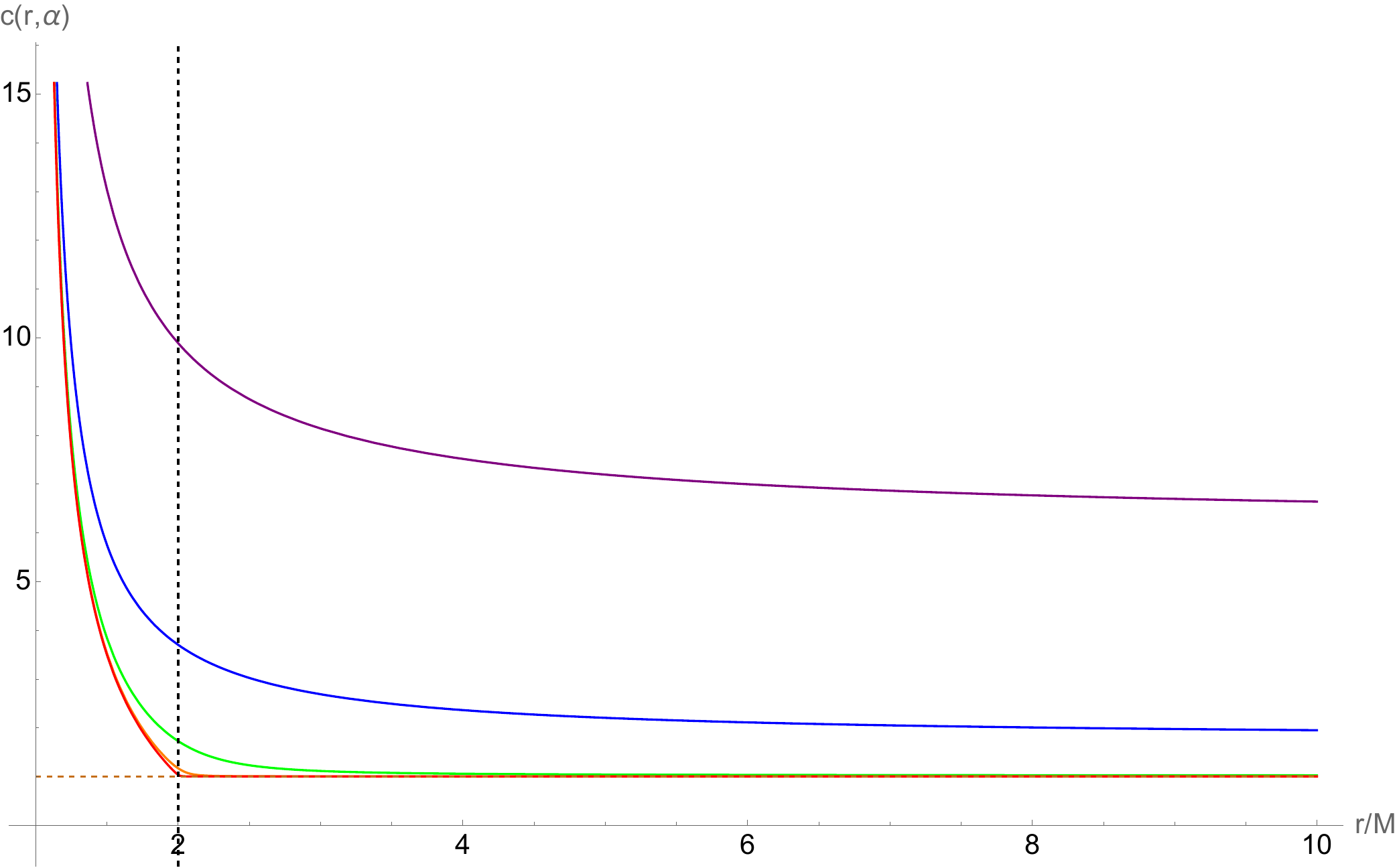}
	\caption{\label{f:cgplot} Group velocity $c_g$ for modes $\psi^{\rm red}_\Omega$ with different $\alpha$: $\a=10$ (purple), $\a=1$ (blue), $\a=10^{-1}$ (green), $\a=10^{-2}$ (orange), $\a= 10^{-3}$ (red). The horizontal dashed line represent the relativistic value $c_g=1$, while the vertical one is located at $r=2M$.}
\end{figure}

As we discussed in the main text, the dynamics of the wave-packet $\psi_\Omega^{\rm red}$ is controlled, in the region far from the UH, by its group velocity $c_g^{\rm red}$. At large radii, and for small $\alpha$, this corresponds to 
\begin{equation}
\label{eq:largerc}
    c_g(r, \alpha)=1+ \frac{3 (r+ 2M)^2 \alpha^2}{2 r^2} + \cdots \,, \quad (r \to \infty, \alpha \to 0) \,,
\end{equation}
which then approaches a constant value in the asymptotic region $c_{g,\infty}(\alpha)=1+3\alpha^2/2$. This $\alpha$-dependent but constant speed, together with fact that $U^\m$ aligns with the time direction for large radii, thus implies that rays arriving to the asymptotic region move following effective light-cones of a disformal metric
\begin{equation}
    \bar{g}_{\mu \nu}=g_{\mu \nu} + (c_g^2-1) U_\mu U_\nu \,.
\end{equation}

Interestingly, this disformal transformation $g_{\m \n} \to \bar{g}_{\mu \nu}$ is an actual symmetry of the EA theory -- together with the replacement $U_\mu \to \bar{U}^\mu = U^\mu / c_g$ \cite{Jacobson10}, and hence physical results cannot depend on it. However, by performing the disformal transformation, we move the position of the Killing horizon to a new location, given by the condition
\begin{equation}
\label{eq:newF}
    0=\Bar{g}_{tt}=F(r,\alpha)= \biggl( 1 - \frac{2M}{r} \biggr) + ( c_{g,\infty}^2(\alpha)-1) \biggl( 1 - \frac{M}{r} \biggr)^2,
\end{equation}
which leads to
\begin{equation}
    r_{\rm H} (\alpha) = \frac{(c_{g,\infty}(\a)+1)}{c_{g,\infty}(\a)}M \,.
\end{equation}
Note that since rays are super-luminal for $\alpha>0$, we find $r_{\rm H}(\alpha)<2M$.

Just looking at Fig.\ref{f:cgplot}, we see immediately that the group velocity, after the KH, assumes a very flat behaviour for low $\a$'s, being almost constant. Indeed from the expansion \eqref{eq:largerc} we see that the $r$-dependence of $c_g$ shows up at order $\alpha^2$. In a very heuristic point of view, the asymptotic observer sees $\psi_\Omega^{\rm red}$ reaching the asymptotic region travelling with an almost constant velocity $c_{g,\infty}(\alpha)$. Thus, they can be interpreted effectively as having been emitted by the Killing horizon of $\bar g_{\m\n}$ in the general relativistic way. Hence we can assign a surface gravity and a temperature to this new Killing horizon, with value
\begin{align}\label{eq:correctionT}
    \kappa_{\rm H}(\alpha)= \frac{1}{2} \frac{{\rm d} F(r,\alpha)}{{\rm d}r} \biggl|_{r_{\rm H} (\alpha)} = \frac{1}{(4 - 12 \alpha^2)M} \quad \mbox{and} \quad T_\alpha = \frac{T_{\rm H}}{1-3\alpha^2} \,,
\end{align}
which agrees with \eqref{eq:temperature_KH}.

\subsection{Effective temperature function}\label{a:ETF}

As a second alternative derivation of  \eqref{eq:temperature_KH}, we introduce now the concept of ``Effective Temperature Function" (ETF). Defined in \cite{Barcelo10}, the ETF measures the degree of peeling of a ray with respect to a given surface. This notion is particularly useful in the case of quasi-horizons \cite{Barcelo10}, and in situations in which event horizons are not yet formed, but the situation is close enough to its final state so that most of the dynamics will approach the latter under certain conditions.

The ETF is defined through the relation $U_+=p^{-1}(U_-)$, which relates the light-cone coordinates at $\mathscr{I}^+$ and $\mathscr{I}^-$, thus connecting the choice of the vacuum state with the coordinate followed locally by the modes. The ETF is then defined as~\cite{Barcelo10}
\begin{equation}\label{eq:ETF}
    \kappa( U_+):= -\frac{{\rm d}^2 U_-}{{\rm d} U_+^2} \left(\frac{{\rm d} U_-}{{\rm d} U_+}\right)^{-1} = -\frac{\Ddot{p}(U_+)}{\Dot{p}(U_+)}.
\end{equation}

In the particular case where $p(U_+)$ takes an exponential form, this captures exactly the peeling surface gravity $\kappa(U_+)=\kappa_{\rm peeling}$. In a general case instead, and even if the peeling behaviour is not perfectly exponential, one can start having Hawking radiation if the variation of $\kappa$ remains adiabatic \cite{Barcelo10}
\begin{equation}
  \biggr| \frac{\Dot{\kappa}(U_+)} {\kappa(U_+)^2}\biggr| \ll 1 
\end{equation}
which implies the approximated constancy of $\kappa(U_+)$ over the time scale associated with the typical period of Hawking quanta -- since the peak frequency of the spectrum is $\omega_{\rm peak}\sim \kappa (U_+)$.

Let us then apply this idea to the case discussed throughout this paper, by treating the KH as a quasi-horizon, following \cite{Barbado12}. It is not an event horizon for rays of arbitrary $\alpha$, but satisfies the previous conditions, and thus rays peel off it, for small $\alpha$. The role of the light-cone coordinates $U_+$ and $U_-$ is played here respectively by the adapted null coordinate for the outgoing (red) modes $\bar u$ and its equivalent one for in-going ones $\bar U$ (blue mode). Note that at $\mathscr{I}^-$ one has $\bar U=U_-$. 

\begin{figure}
	\includegraphics[scale=0.35]{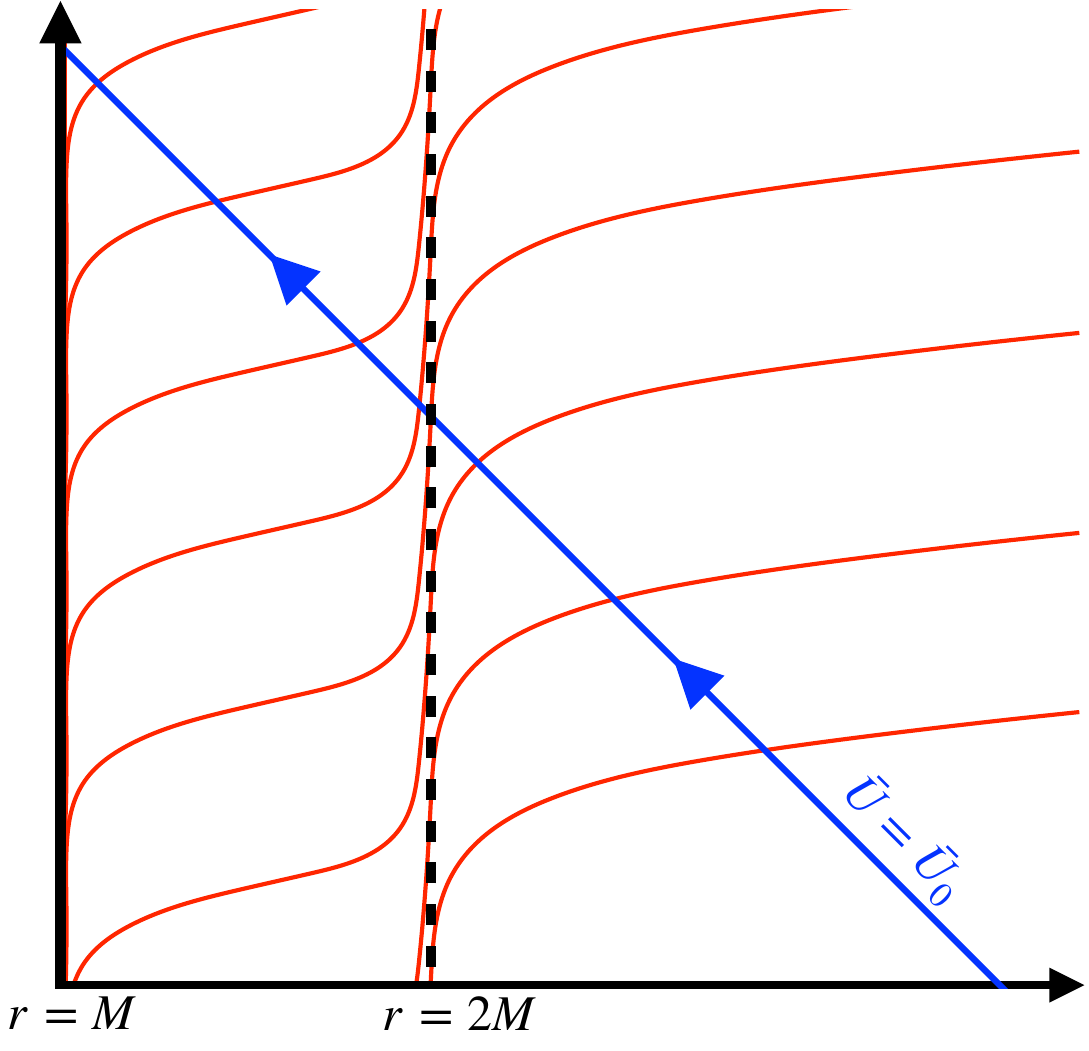}
	\caption{\label{f:etf} Constant $\Bar{U}=\Bar{U}_0$ observer (in blue) which crosses a congruence of $\Bar{u}= \rm const.$ lines (in red).}
\end{figure}

Let us now consider the variation of $\bar u$ as perceived by an infalling observer along a blue mode trajectory. We take such an observer to have Killing energy well below $\Lambda$ at $\mathscr{I}^-$, so it will be approximately relativistic all along its path while infalling into the KH. Indeed, for low-energy blue modes, for which $k_{\rm blue}(r,\alpha) \ll 1$, we have that the trajectory can be described by a parameter $\lambda$ as
\begin{equation}\label{eq:bluemodetraj}
    \frac{{\rm d}t}{{\rm d} \lambda} = - \frac{r}{r-2M} \biggl[ 1 + \frac{3}{2} \frac{k_{\rm blue}^2}{\Lambda^2} \biggr(\frac{2M}{r}-1 \biggr) \biggr] \frac{{\rm d}r}{{\rm d} \lambda} \,,
\end{equation}
which near the KH coincides exactly with the relativistic null observer along ${\rm d} v=0$. Since we will be interested in that region, in the following we will neglect the subleading term in \eqref{eq:bluemodetraj}. Then, using eq.\eqref{eq:bluemodetraj} and \eqref{eq:dubar}, we can label a constant $\bar U$ line with $\Bar{u}$ getting
\begin{equation}
\label{eq:labelubar}
    \frac{{\rm d} \Bar{u}}{{\rm d} \lambda} = \frac{ {\rm d}t}{{\rm d} \lambda} + \frac{c_g^{\rm red}(r,\alpha) U_r + S_r}{c_g^{\rm red}(r, \alpha) U_t + S_t}  \frac{{\rm d}r}{{\rm d} \lambda}= \biggl( \frac{c_g^{\rm red}(r,\alpha) U_r + S_r}{c_g^{\rm red}(r, \alpha) U_t + S_t}  - \frac{r}{r-2M} \biggr) \frac{{\rm d}r}{{\rm d} \lambda} \,.
\end{equation}
This relation describes exactly the situation depicted in Fig.\ref{f:etf}, corresponding to an observer travelling along a $\bar U=\bar U_0= \rm const.$ line crossing the outgoing red trajectories.

Let us now make a convenient choice that simplifies the computation, i.e.~we label points along a $\bar U$=const trajectory using their radius (this is always allowed as long as the relation between $\lambda$ and $r$ is monotonic, as in our case).   This is tantamount to choosing a parameter $\lambda$ so that coinciding with the radial coordinate.
In this case we have $\partial_\lambda r=1$ and, expanding \eqref{eq:labelubar} using \eqref{eq:largerc}, we obtain
\begin{equation}
   \frac{{\rm d} \Bar{u}}{{\rm d} \bar U}=\frac{{\rm d} \Bar{u}}{{\rm d} r}= -\frac{2 r}{r-2 M}+\frac{3 \alpha ^2 \left(4 M^2+4 M r+r^2\right)}{2 (r-2 M)^2}+O\left(\alpha ^3\right) \,,
\end{equation}
that we can integrate to
\begin{equation}
\label{eq:u(U)}
  \Bar{u}(r)= \frac{1}{2} \left(-\frac{48 \alpha ^2 M^2}{r-2 M}+8M \left(3 \alpha ^2 -1 \right) \log (r-2 M)+\left(3
   \alpha ^2-4\right) r\right) \,.
\end{equation}
Neglecting the terms which are finite for $r \to 2M$, we can invert the expression above getting finally the relation ${\bar U}(\Bar{u})$
\begin{equation}
  {\bar U}(\Bar{u})=r(\Bar{u})=2M +M \frac{6 \alpha^2}{3 \alpha^2 -1} \frac{1}{W \biggl(\frac{6 \alpha^2}{3 \alpha^2 -1} e^{- \frac{1}{4 M (3 \alpha^2 -1)} \Bar{u}}\biggr)} \,,
\end{equation}
where the function $W(x)$ is the (principal branch of the) Lambert function\footnote{The Lambert function is defined as the solution of the equation $W(x) e^{W(x)}=x$.}. Let us notice that, when $\alpha \to 0^+$, the expression for $\bar{U}(\Bar{u})$ matches, as expected, the one for the Kruskal–Szekeres null coordinates plus ${\cal O}(\alpha^2)$ corrections, 
\begin{equation}
 \lim_{\alpha\to 0^+}  {\bar U}(\Bar{u})=M e^{- \Bar{u}/4M} \biggl( 1 - 3 \a^2 \frac{\Bar{u}}{4M} + \mathcal O(\a^4) \biggr) \,,
\end{equation}
thus recovering exactly the relativistic peeling with $\kappa=(4M)^{-1}$. 

With these ingredients, we are ready to compute the ETF and expand it at small $\alpha$
\begin{equation}
\label{eq:effectivek}
  \kappa (\Bar{u})=  -\frac{{\rm d}^2 {\bar U}}{{\rm d} \Bar{u}^2} \left(\frac{{\rm d} {\bar U}}{{\rm d} \Bar{u}}\right)^{-1} =\frac{1}{4M (1-3 \alpha^2 )}\frac{1 + 2 W \biggl(\frac{6 \alpha^2}{3 \alpha^2 -1} e^{- \frac{1}{4 M (3 \alpha^2 -1)} \Bar{u}}\biggr)}{\biggl[ 1+ W \biggl(\frac{6 \alpha^2}{3 \alpha^2 -1} e^{- \frac{1}{4 M (3 \alpha^2 -1)} \Bar{u}} \biggr) \biggr]^2} =  \frac{1}{4M}(1 + 3 \alpha^2 + \cdots) \,,
\end{equation}
which is constant up to corrections of order ${\cal O}(\a^4)$, hence automatically satisfying the adiabatic condition. The result obtained in \eqref{eq:effectivek} leads to the same temperature in \eqref{eq:temperature_KH}.

As a final interesting observation, let us note that the same result displayed here can be obtained by taking the limit in which the observer approaches the KH. In \eqref{eq:u(U)}, this corresponds in $\Bar{u}$ to the limit $\Bar{u} \to - \infty$. In the ETF, this corresponds to
\begin{equation}
  \lim_{\Bar{u} \to - \infty} \kappa(\Bar{u})= \frac{1}{4M(1-3\alpha^2)},
\end{equation}
which can be interpreted as the rays departing the KH with a constant $\a$-dependent exponent.

\section{Perturbative analysis of the hard branches}
\label{a:pert_hard}

Here we derive the behavior of the red modes in the intermediate region in between the UH and the KH, used in \eqref{eq:wkb}. See also \cite{Schutzhold13}. 

Let us start from eq.\eqref{eq:krho_eq}. Taking $n=2$, we get, after substituting $\Lambda=\Omega/\a$
\begin{align}
    \frac{\a^2}{\Omega^{2}} k^{4} + \biggr( 1- \frac{V^2}{N^2} \biggr) k^2 + \frac{2 \Omega V}{N^2} k - \frac{\Omega^2}{N^2} =0\,.
\end{align}
We remark that within the region of interest, $N>0$.

Fixing the value of $\Omega>0$, the previous equation displays two different behaviors with respect to $\alpha$, as it can be recognised by the fact that $\alpha^2$ multiplies the highest power of $k$. The first possibility corresponds to a regular solution when $\alpha \rightarrow 0$, which matches \eqref{eq:krho_expansion_sol}; the other option takes the form
\begin{align}
    k= \frac{f(r)}{\a}+ O(1)\,,
\end{align}
where $f(r)$ is determined from 
\begin{align}
  f(r)^2 \biggl[ \frac{f(r)^2}{\Omega^{2}} + \biggr( 1- \frac{V^2}{N^2} \biggr) \biggr]=0\,.
\end{align}

Besides the trivial solution $f(r)=0$, we have
\begin{align}\label{eq:krho_hard_expansion}
  k^\pm = \pm \frac{\Omega}{\a} \sqrt{ \frac{V^2}{N^2}-1}=\pm \frac{\Omega}{\a N} \sqrt{ -|\chi|^2 }\,,
\end{align}
which behaves as $\a^{-1}$, as already anticipated in \eqref{eq:wkb}. 

Let us stress here that, while the solution found in \eqref{eq:krho_expansion_sol} corresponds to the soft branches of $\psi^{\rm red}$ and $\psi^{\rm orange}$ -- respectively outside and inside the KH -- this non-analytic solution in $\a$ characterizes the hard part of these same rays. In fact, we can observe that the solutions in \eqref{eq:krho_hard_expansion} become imaginary when $|\chi|^2>0$, which corresponds to the exterior of the KH.

As a final remark, let us note that the solutions discussed here are the same ones studied in \cite{Schutzhold13}, so our independent analysis coincide. Indeed, the result in \cite{Schutzhold13} is valid within the approximation $\kappa_{\rm KH} \ll \Lambda$. In our case, the radiation which reaches the KH is peaked at $\Omega = \kappa_{\rm UH}/ \pi = 4  \kappa_{\rm KH}/ \pi$. Their constraint then translates to $\Omega \ll \Lambda$, so to $\a \ll 1$, hereby matching again our results.


\section{Inner product and normalization}\label{a:inner_product}

In the main text we discussed the normalization of the modes, but we did not introduce the inner product under which that normalization takes effect. Given two solutions of the equations of motion, we thus define
\begin{align}\label{eq:inner_product}
 \langle \phi_1,\phi_2 \rangle_\tau = -i\int_{\Sigma_\tau}\!\!\!{\rm d}^3x\sqrt{-\gamma} \;U^\mu ( \phi_1 \nabla_\mu \phi_2^* - \phi_2^* \nabla_\mu \phi_1 )\,.
\end{align}
where the integral is evaluated at a constant-time surface $\Sigma_\tau$ such that $U_\mu {\rm d}x^\mu |_{\Sigma_\tau}=0$, meaning that $\Sigma_\tau$ is a leaf of the preferred foliation. Using the equations of motion, it is easy to show that this definition is $\tau-$independent, so we will omit the index $\tau$ from now on.

This product can now be used to normalize the modes in the standard way, thus introducing also the concept of positive and negative norm modes. In Section \ref{s:QS} we used this to impose a specific normalization of the modes. In the case of monochromatic waves, for which $\langle \phi_{\Omega_1}, \phi_{\Omega_2} \rangle = \delta_{\Omega_1, \Omega_2}$, this leads to
\begin{align}\label{eq:inner_product}
\phi_\Omega = \frac{1}{\sqrt{4 \pi \omega}}e^{-i \int (\omega U_\m {\rm d}x^\m + k S_\m {\rm d}x^\m )}\,,
\end{align}
where the amplitude is thus constant within the eikonal approximation. 

For the wave-packet $\psi^{\rm red}_\Omega$ we may require as well $\langle \psi^{\rm red}_{\Omega_1}, \psi^{\rm red}_{\Omega_2} \rangle = \delta_{\Omega_1, \Omega_2}$. The amplitude of the packet -- see \cite{Parentani15} -- then reads:
\begin{align}
 \psi^{\rm red}_\Omega = \sqrt{ - \frac{V c_g^{\rm red}}{16 \pi^2 \omega}} e^{i \mathcal{S}_0}\,,
\end{align}
where $\mathcal{S}_0$ is given by \eqref{eq:pp_action_general}.

\bibliography{bibliography.bib}{}
\end{document}